\documentclass{pasa}%
\usepackage{bigdelim}
\usepackage{graphicx}
\usepackage{lscape}
\usepackage{longtable}
\usepackage{todonotes}
\usepackage{rotating}
\usepackage{threeparttable}
\usepackage{hyperref}
\usepackage{multirow}
\usepackage{subcaption}

\def\nar{\ref@jnl{New A Rev.}}          

\title[POGS I: Initial Results]{The POlarised GLEAM Survey (POGS) I: First Results from a Low-Frequency Radio Linear Polarisation Survey of the Southern Sky}

\author[Riseley et al.]{C.~J.~Riseley$^1$\thanks{Corresponding author email: \url{chris.riseley@csiro.au}}, \hspace{0.001cm} E.~Lenc$^{2,3}$, C.~L.~Van~Eck$^{4,5}$, G.~Heald$^{1}$, B.~M.~Gaensler$^{5}$, C.~S.~Anderson$^{1}$, P.~J.~Hancock$^{6}$, N.~Hurley-Walker$^{6}$, S.~S.~Sridhar$^{7}$, and S.~V.~White$^{6}$
\affil{$^1$ CSIRO Astronomy and Space Science, PO Box 1130, Bentley, WA 6102, Australia}
\affil{$^2$ CSIRO Astronomy and Space Science, PO Box 76, Epping, NSW 1710, Australia}
\affil{$^3$ School of Physics, Sydney Institute for Astronomy, The University of Sydney, NSW 2006, Australia}
\affil{$^4$ Department of Physics and Astronomy, University of Calgary, Calgary, Alberta, T2N 1N4, Canada}
\affil{$^5$ Dunlap Institute for Astronomy and Astrophysics, 50 St. George St, University
of Toronto, ON M5S 3H4, Canada}
\affil{$^6$ International Centre for Radio Astronomy Research, Curtin University, Bentley, WA 6102, Australia }
\affil{$^7$ Netherlands Institute for Radio Astronomy (ASTRON), Postbus 2, 7990 AA, Dwingeloo, The Netherlands}
}

\jid{PASA}
\doi{10.1017/pas.\the\year.xxx}
\jyear{\the\year}

\usepackage{aas_macros}
\usepackage{hyperref} 
\hypersetup{colorlinks,citecolor=blue,linkcolor=blue,urlcolor=blue}

\begin{document}

\begin{frontmatter}
\maketitle

\begin{abstract}
The low-frequency polarisation properties of radio sources are poorly studied, particularly in statistical samples. However, the new generation of low-frequency telescopes, such as the Murchison Widefield Array (MWA; the precursor for the low-frequency component of the Square Kilometre Array) offers an opportunity to probe the physics of radio sources at very low radio frequencies. In this paper, we present a catalogue of linearly-polarised sources detected at 216~MHz, using data from the Galactic and Extragalactic All-sky MWA (GLEAM) survey. Our catalogue covers the Declination range $-17\degree$ to $-37\degree$ and 24 hours in Right Ascension, at a resolution of around 3~arcminutes. We detect 81 sources (including both a known pulsar and new pulsar candidate) with linearly-polarised flux densities in excess of 18~mJy across a survey area of approximately 6400 square degrees, corresponding to a surface density of 1 source per 79 square degrees. The level of Faraday rotation measured for our sources is broadly consistent with those recovered at higher frequencies, with typically more than an order of magnitude improvement in the uncertainty compared to higher-frequency measurements. However, our catalogue is likely incomplete at low Faraday rotation measures, due to our practice of excluding sources in the region where instrumental leakage appears. The majority of sources exhibit significant depolarisation compared to higher frequencies; however, a small sub-sample repolarise at 216~MHz. We also discuss the polarisation properties of four nearby, large-angular-scale radio galaxies, with a particular focus on the giant radio galaxy ESO~422$-$G028, in order to explain the striking differences in polarised morphology between 216~MHz and 1.4~GHz. 

\end{abstract}

\end{frontmatter}

\section{INTRODUCTION }
\label{sec:intro}
Investigating the origins of cosmic magnetism is one of the key science drivers behind the Square Kilometre Array \citep[SKA; see][for details on probing cosmic magnetism with the SKA]{Gaensler2004,JohnstonHollitt2015}. One means by which observers hope to discriminate between models of the origins of cosmic magnetism is through construction of a grid of polarised sources \citep{Beck2004} that can be used to statistically probe magnetic fields in the under-dense regions of the Universe (the filamentary and void regions of the large-scale structure).

Polarimetry at low radio frequencies has historically been challenging for a number of reasons. These include Faraday depolarisation (whereby polarised sources depolarize with increasing wavelength, e.g. \citealt{Burn1966,Farnsworth2011,Anderson2015}), beam depolarisation (where changes in polarisation angle within a given beam result in an apparent loss of polarised signal), ionospheric depolarisation (where ionospheric effects induce an additional Faraday rotation, and can totally decorrelate the observed polarised signal) and also the poor sensitivity of many historic low-frequency instruments. However, with the new generation of low-frequency telescopes such as the LOw-Frequency ARray \citep[LOFAR;][]{vanHaarlem2013} and Murchison Widefield Array \citep[MWA;][]{Tingay2013} and recent advances in methodology, low-frequency polarimetry is experiencing a renaissance, with a wealth of recent scientific results being generated from these instruments \citep[e.g.][]{Mulcahy2014,Jelic2015,Lenc2016,Lynch2017,VanEck2017,VanEck2018,Lenc2018}. 

Where linearly-polarised radio emission encounters a magnetic field during propagation from source to observer, the plane of polarised emission undergoes Faraday rotation, according to
\begin{equation}\label{eq:chi}
    \chi(\lambda^2) = \chi_0 + \phi \lambda^2 = \chi_0 + 0.81 \int_{\rm{LOS}} n_{\rm{e}} {\textbf{B}}\cdot {\rm{d}}{\boldsymbol{\ell}}
\end{equation}
where $\chi(\lambda^2)$ is the observed polarisation angle, $\lambda$ is the observing wavelength, $\chi_0$ is the intrinsic source polarisation angle, and $\phi$ defines the Faraday depth (in rad m$^{-2}$). The integral term in Equation~\ref{eq:chi} indicates that the Faraday depth depends on the number density of free electrons $(n_{\rm{e}})$ and the magnetic field component along the line-of-sight (LOS; ${\textbf{B}}\cdot {\rm{d}} {\boldsymbol{\ell}})$ from the source to the observer.

In the literature, "Faraday depth" (FD) and "rotation measure" (RM) are often used interchangeably. However, strictly speaking, \cite{Brentjens2005} and \cite{Burn1966} define FD and RM as two distinct quantities: FD is the more generalised integral quantity, and RM is defined as the slope of polarisation angle $\chi$ versus $\lambda^2$, i.e. 
\begin{equation}\label{eq:rm}
    {\rm{RM}} = \frac{ {\rm{d}}\chi(\lambda^2)}{{\rm{d}}\lambda^2}
\end{equation}
In this paper, we will use RM when referring to literature that uses the term \citep[primarily][]{Taylor2009} and FD at all other times. Under the assumption that a medium along the LOS is solely Faraday-rotating \citep[rather than emitting as well as rotating, see e.g.][]{Brentjens2005,Heald2009} the two can be equated.

The GaLactic and Extragalactic All-sky MWA survey \citep[GLEAM;][]{Wayth2015} covers the entire sky south of Declination $+30\degree$. The GLEAM extragalactic catalogue \citep{HurleyWalker2017} covers 24831 square degrees of sky below this Declination and at Galactic latitudes $|b| \geq 10\degree$. With this excellent sky coverage, and recent advances in techniques for low-frequency polarimetry calibration using the MWA \citep{Lenc2017} as well as source-finding and verification techniques \citep[e.g.][]{Farnes2018,VanEck2018} we can attempt to investigate the distribution and properties of polarised sources at low radio frequencies in the pre-SKA era.

In this paper, we present the first results from the POlarisation from the GLEAM Survey (POGS) project. This is a new large-scale reprocessing of visibilities from the GLEAM survey, which aims to catalogue and characterise the low-frequency linearly-polarised source population. We discuss the methods we have used to mitigate the various effects that have limited such studies in the past, and present our initial catalogue. We also discuss some of the physics that can be inferred from the polarisation properties of the sources we detect. Lastly, we discuss future directions for POGS.

Throughout this paper, we assume a $\Lambda$CDM cosmology of $H_0=67.8~{\rm{km~s}}^{-1}~{\rm{Mpc}}^{-1}$, $\Omega_{\rm{m}} = 0.308$, $\Omega_{\Lambda} = 0.692$ \citep{Planck2016}. All errors are quoted to $1\sigma$, and we adopt the spectral index convention that $S\propto \nu^{\alpha}$.

\section{Sample selection and data processing}\label{sec:processing}
For this first paper, in which we verify our methods and results, we selected a sub-sample from the full GLEAM survey: the drift scan strip centred on Declination $-27\degree$, as this strip passes through zenith, where instrumental leakage should be reduced (see the discussion in the next section). Additionally, selecting this Declination strip allows cross-comparison with the NRAO VLA Sky Survey \citep[NVSS;][]{Condon1998} RM catalogue \citep{Taylor2009}. 

The Dec.~$-27\degree$ strip was covered during four observing runs (between 2013 August and 2014 June) which overlap by approx. $2-4$h in Right Ascension. We have reprocessed all four epochs, covering the full 24 hours in RA, between $-37\degree \lesssim \delta \lesssim -17\degree$. Given the instantaneous MWA bandwidth of 30.72~MHz, the GLEAM observations were divided into five bands covering the range $72-231~{\rm{MHz}}$. On each night's observing of a given Declination strip and RA range, the frequencies were cycled over the course of ten minutes \citep[see][]{Wayth2015}. Additionally, we selected the top GLEAM band $(200-231~{\rm{MHz}})$ as this retains greatest sensitivity to large Faraday depths. Note that we discuss future directions for POGS in Section~\ref{sec:future}. We present the observation dates and Right Ascension ranges covered in Table~\ref{tab:obs}.

Calibration and imaging were performed using the Real-Time System \citep[RTS;][]{Mitchell2008}. Whilst a number of calibrators were observed during each run, we selected the calibrator source observed at the highest elevation to transfer solutions to our drift-scan snapshots. This was done to both i) ensure greatest possible signal-to-noise ratio (SNR) on our calibrator and ii) mitigate uncertainties caused by differences in the primary beam between calibrator scans and drift scans.

\begin{table}[!ht]
\caption{Observing details for our sub-sample of GLEAM observations.}
\centering
\begin{tabular}{@{}cc@{}}
\hline\hline
Right Ascension range & Obs Date \\
(J2000) & \\
\hline%
 19 -- 3.4h & 2013-08-10 \\
 00 -- 08h  & 2013-11-25 \\
 06 -- 16h  & 2014-03-03 \\
 12 -- 22h  & 2014-06-09 \\ 
\hline\hline
\end{tabular}
\label{tab:obs}
\end{table}

Unlike many conventional imaging algorithms, the RTS does not perform deconvolution without a-priori knowledge of the source population. Whilst this is not critical for polarimetry, where we are not limited by source confusion, deconvolution can help mitigate sidelobes from brighter Stokes I sources in the field. For this work, we imaged a $20\times20$ square degree region around the phase centre for each snapshot, using the native GLEAM frequency resolution of 40~kHz. Whilst this limits our per-channel image sensitivity, we can use the RM synthesis technique \citep[e.g.][]{Burn1966,Brentjens2005} to recover even low-level polarised emission that might be otherwise remain undetected. In this project, we are interested in cataloguing extragalactic objects, so we wish to minimise contamination from the polarized Galactic foreground \citep[e.g.][]{Bernardi2013,Lenc2016}. As such, we employed an inner $uv$-cut of $50~\lambda$; we also employed an outer $uv$-cut of $1~{\rm{k}}\lambda$ in order to maintain near-constant resolution across the entire band. We also employed a \texttt{robust}$=-1$ weighting scheme \citep[identical to that used for the GLEAM survey;][]{HurleyWalker2017}. As a result, our final resolution is around $190\times170$~arcsec. Our pixel size was 40~arcsec.

\subsection{Correcting for instrumental leakage}
Instrumental leakage results from Stokes I signal `leaking' into other Stokes parameters, causing apparent polarised signal. This leakage is typically caused by errors in the primary beam model, and is most evident in Stokes Q (as both Q and I are formed from the same correlation products for linear feeds). From early MWA observations in the mid-frequency band (centred on 154~MHz) and near zenith $(-27\degree)$, the leakage was found to be of the order of the order of 1 per cent toward the field centre, and around 4 per cent toward the periphery \citep{Bernardi2013}. However, in the higher frequency bands and away from zenith, the instrumental leakage can be as high as 40 per cent \citep[e.g.][]{Lenc2017}. 

Despite improvements in the MWA primary beam model \citep{Sutinjo2015}, residual imperfections in both the models and the MWA dipole antennas, as well as any failures during the observations result in leakage of Stokes I signal into the other polarisation components. Recently, \cite{Lenc2017} presented an empirical method with which to correct for instrumental leakage, that exploits the large field-of-view (FOV) and well-sampled instantaneous \emph{uv}-coverage of the MWA. This method is also discussed in detail by \cite{Lenc2018}.

\begin{figure*}
\begin{center}
\includegraphics[width=0.95\textwidth]{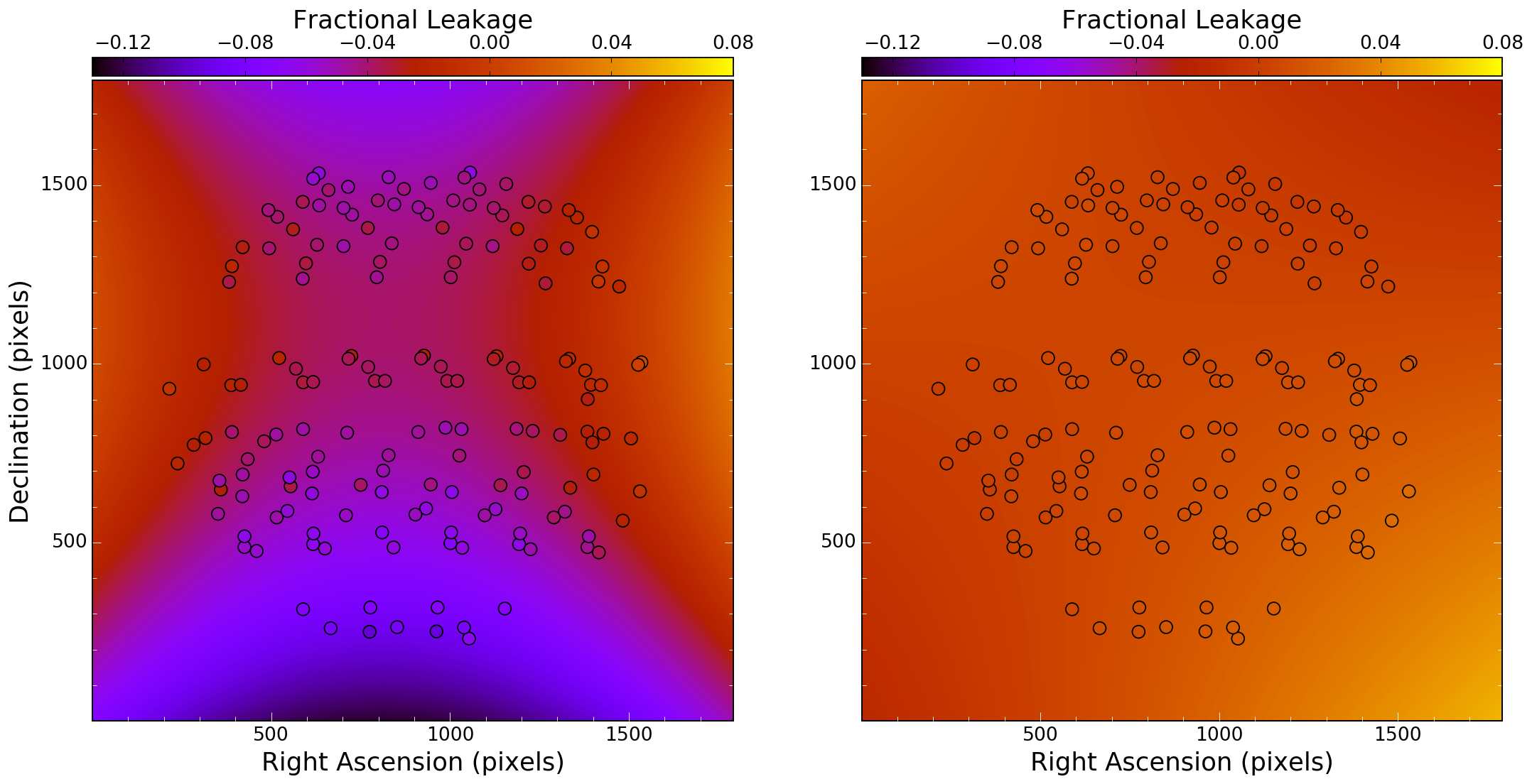}
\caption{Example point-source leakage fit results over the observed FOV for the GLEAM-1.4 drift scans in the 204-232 MHz band. Compact sources are marked by circles, with colour scales matched to the leakage surface fit (background colourscale). Fits were performed independently for Stokes Q (left panel) and Stokes U (right panel). Note that negative leakage means that an unpolarised source would have apparent negative polarised flux density.}\label{fig:leakage}
\end{center}
\end{figure*}

We refer the reader to \cite{Lenc2017,Lenc2018} for full details, but in short, this method uses the snapshot observations of sources that drift through the MWA primary beam. Assuming all sources are unpolarised in the continuum images, the spatial variation of the leakage can be fitted for, deriving a (frequency-independent) `leakage surface'. This is then used to perform an image-plane subtraction of the Stokes I leakage from each snapshot. We present the fitted leakage surfaces for Stokes Q and U from the $12-22$h RA GLEAM observing run in Figure~\ref{fig:leakage}.

\begin{figure*}
\begin{center}
\includegraphics[width=0.95\textwidth]{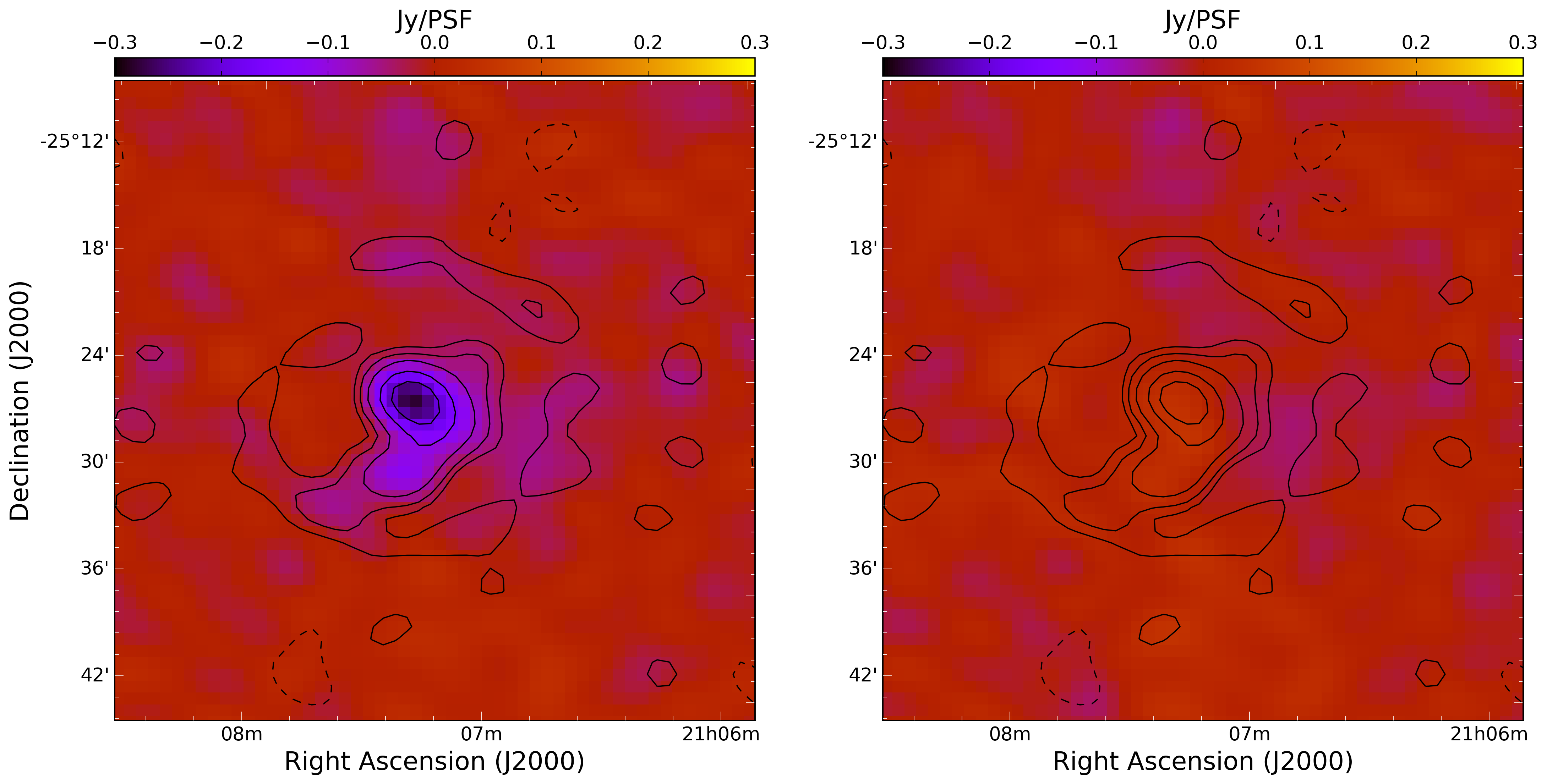}
\caption{The effect of leakage correction on Stokes Q continuum images of a bright $($peak Stokes I flux density $\sim 12~{\rm{Jy~PSF}}^{-1})$ source. \emph{Left panel:} prior to instrumental correction. \emph{Right panel:} after correction. Contours are dirty Stokes I continuum in the 216~MHz band, at $[-1,1,2,4,8,12]$ Jy PSF$^{-1}$. Panels are set to matching colour scales. Following correction, the peak residual is $0.03$~Jy PSF$^{-1}$ (or $P/I = 0.26\%$).}\label{fig:leakage_cor}
\end{center}
\end{figure*}

From Figure~\ref{fig:leakage}, the typical leakage is of the order of $2-4$ per cent toward the centre of the beam in Stokes Q, rising to around 10 per cent toward the edge of the beam. The Stokes U leakage is significantly less, typically around the 1 per cent level near the beam centre and up to 3 per cent in the periphery. Figure~\ref{fig:leakage_cor} shows the effect of leakage correction on the Stokes Q continuum image of a bright source, centred on the source GLEAM J210722$-$252556 (apparent peak flux density approximately 12~Jy PSF$^{-1}$). Prior to the correction, the measured Stokes Q flux density is around $-0.26$~Jy PSF$^{-1}$, suggesting the leakage of the order of $2.2$ per cent; following leakage correction, the Stokes Q flux density is significantly reduced at $0.03$~Jy PSF$^{-1}$ (or $0.26$ per cent leakage). Typically, the residual leakage seen in the FD spectra of our sources is $\lesssim0.5\%$. Note that this source exhibits no real polarised emission.

\subsection{Correcting for ionospheric Faraday rotation}
For the new generation of low-frequency telescopes - possessing large fractional bandwidths - the precision with which observers can measure Faraday depths is greater than at higher frequencies (see Equation~\ref{eq:rmparms1}). Consequently, ionospheric Faraday rotation can become the dominant source of uncertainty in measured Faraday depths. 

Broadly-speaking, the relatively short integration time of an individual GLEAM drift scan (approx. 2 minutes) means that ionospheric effects should be constant over an individual snapshot. However, in practice, it was found that the ionosphere could become highly disturbed on short timescales \citep[e.g.][]{Loi2015}. These observations were flagged and discarded. Given that even a relatively small variation in ionospheric effects can depolarize the brightest polarised sources \citep[e.g.][]{Lenc2017} however, we must apply corrections for ionospheric Faraday rotation.

We used the RMextract tool\footnote{\url{https://github.com/lofar-astron/RMextract}} \citep{Mevius2018}, which uses maps of the total electron content (TEC) derived from GPS data, in conjunction with models of the Earth's magnetic field to provide a correction for the total electron content and ionospheric rotation measure. Whilst the spatial and temporal sampling is moderate, the overall correction is sufficient, given the short integration time of the drift-scan snapshots and the compact nature of the MWA. The ionospheric RMs reported by RMextract (as a function of UT) were then used to de-rotate the Stokes Q and U image cubes. The typical ionospheric RM correction applied was of the order of $-1$ to $-6$ rad m$^{-2}$.

\subsection{Mosaicking}
The GLEAM observing strategy made use of drift scans, resulting in significant overlap between fields. In order to improve our sensitivity, we mosaicked our leakage- and ionospheric-RM-corrected $Q$ and $U$ snapshot images to form a single mosaic for each observing run. We employed \textsc{SWarp} \citep{Bertin2002} to re-project and mosaic all snapshots on a per-channel basis, producing approx. 650 mosaics for each observing run, weighted according to the square of the primary beam. The resulting mosaics were then stacked, forming the $Q(\lambda)$ and $U(\lambda)$ image cubes used as the inputs for RM synthesis. 

Note that the significant spatial oversampling between snapshots means that each location in our mosaicked cubes will effectively have a pseudo weighted average ionospheric FD correction applied. As such, we cannot state the exact ionospheric FD that was subtracted for a given source. For a handful of bright polarised sources in the region covered in this paper, we verified empirically that this correction yielded consistent FD peaks between individual snapshots.

\subsection{Rotation Measure synthesis}
Subsequently, we performed RM synthesis \citep[e.g.][]{Brentjens2005,Heald2009}. The relevant parameters for RM synthesis -- the Faraday-space resolution $(\Delta \phi)$, maximum Faraday depth $(|\phi_{\rm{max}}|)$ and maximum scale in Faraday space $(\rm{max.~scale})$ -- are defined as 
\begin{subequations}
	\begin{equation}
    	\Delta \phi = 2 \sqrt{3} / \Delta (\lambda^2) \label{eq:rmparms1}
	\end{equation}
	\begin{equation}
    	{\rm{max.~scale}} = \pi/\lambda_{\rm{min}}^2 \label{eq:rmparms3}
	\end{equation}
	\begin{equation}
	|\phi_{\rm{max}}| = \sqrt{3} / \delta(\lambda^2) \label{eq:rmparms2}
	\end{equation}
\end{subequations}
where $\Delta(\lambda^2)$ is the difference in wavelength-squared across the observing bandwidth, $\delta(\lambda^2)$ is the wavelength-squared difference across each channel and $\lambda_{\rm{min}}$ is the wavelength of the highest-frequency channel. 

\begin{figure}
\begin{center}
\includegraphics[width=0.48\textwidth]{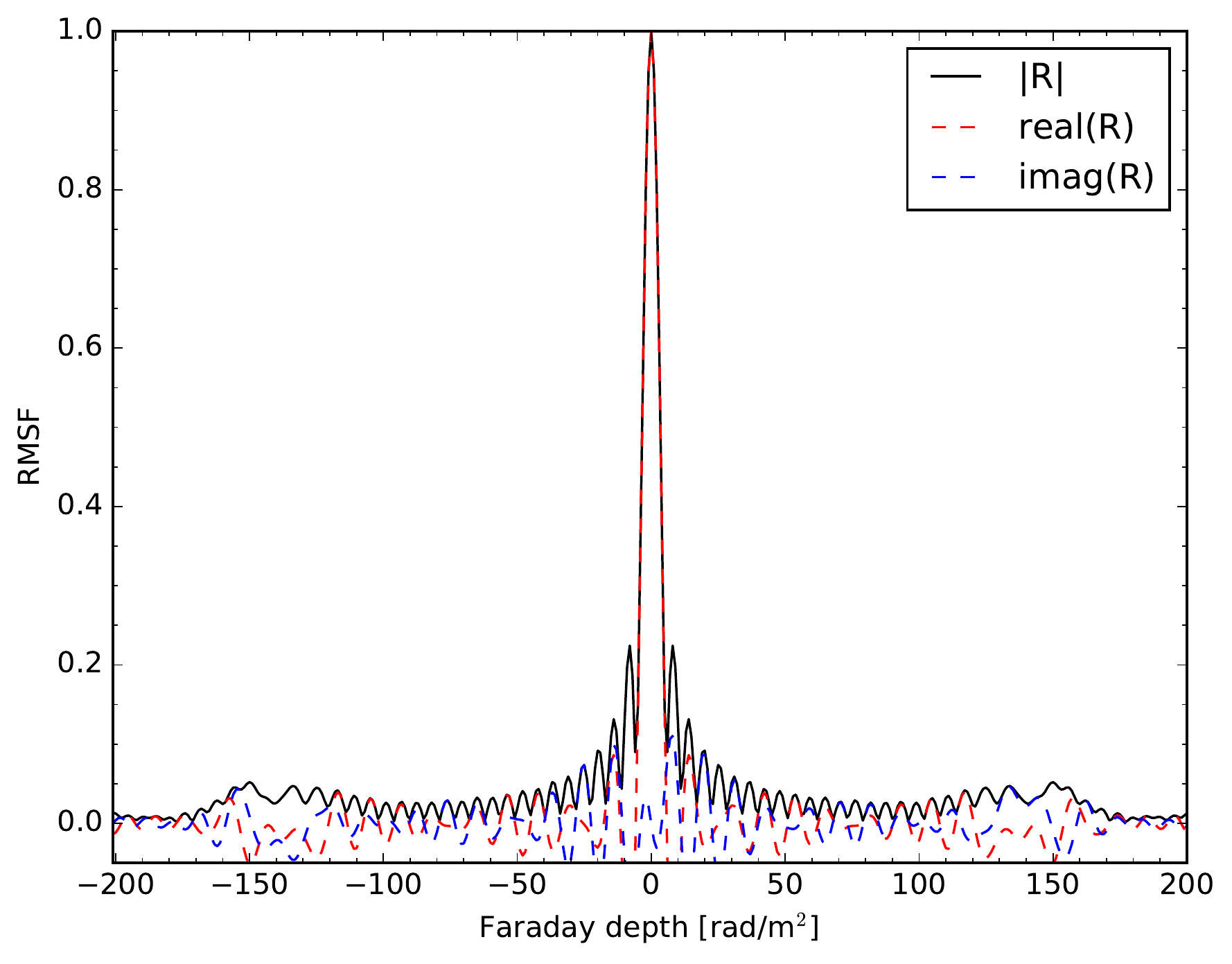}
\caption{An example Rotation Measure Spread Function (RMSF) for a POGS RM cube. Solid line depicts the magnitude of the RMSF (labelled `R'); dashed lines denote the real and imaginary components.}\label{fig:rmsf}
\end{center}
\end{figure}

Our observations cover the frequency range $200.32-231.04$~MHz. From Equation~\ref{eq:rmparms1} $\Delta \phi = 6.23$~rad~m$^{-2}$, meaning we can measure Faraday depths to high precision. However, the maximum scale size we can recover is small: from Equation~\ref{eq:rmparms3}, the maximum scale we can recover is around 1.9~rad~m$^{-2}$. Our use of the native 40~kHz GLEAM channelisation means that we retain sensitivity to large Faraday depths; from Equation~\ref{eq:rmparms2}, $|\phi_{\rm{max}}|=1937$ rad m$^{-2}$. However, due to computational restrictions, for this initial study we have limited our investigation to the FD range $|\phi| \leq 200 $ rad m$^{-2}$. We present the Rotation Measure Spread Function (RMSF) for the MWA in this frequency range in Figure~\ref{fig:rmsf}. Away from the Galactic plane, FD are expected to be $|\phi|\ll200$ \citep[e.g.][]{Taylor2009,Schnitzeler2010}; indeed, only 0.75\% of NVSS sources in the region considered here have RMs outside our FD range. As a result, we do not expect significant bias in our sample.

We used the new GPU-based \texttt{cuFFS} recipe\footnote{\url{https://github.com/sarrvesh/cuFFS}} (Sridhar et al., submitted) to perform RM synthesis on our mosaicked frequency cubes (each of which covers approximately $2400-3000$ square degrees of sky). \texttt{cuFFS} has been optimised for processing large data cubes on GPU-based systems, primarily for processing LOFAR polarimetric data. For details concerning the performance of cuFFS, see Sridhar et al.

\subsection{Noise estimation}\label{sec:noise}
We perform our source-finding in $P(\phi)$-cubes, so we need to characterise the linear polarisation noise. The noise in our FD cubes is expected to follow a Rayleigh distribution in $P$ (see for example \citealt{Macquart2012,Hales2012}; or for the more general case of Ricean statistics, see \citealt{Wardle1974}). In order to estimate our FD cube noise, we followed the same method described by \cite{VanEck2018}. We fitted a Rayleigh distribution to the histogram of polarised flux densities (as a function of Faraday depth) at the location of each pixel. Following visual inspection of our FD cubes, we excluded the range $|\phi|\leq 20$ rad m$^{-2}$ from this estimate, as this region encapsulated both residual instrumental leakage and sparsely-sampled Galactic foreground. 

The typical noise level away from the field edges in our FD cubes was of the order of $3-5$~mJy PSF$^{-1}$ RMSF$^{-1}$. As a result of the GLEAM drift scan observing mode, the bulk of each mosaic is spatially oversampled by a factor $\gtrsim10$. From \cite{Franzen2016}, the uniformly-weighted Stokes $V$ noise for a single snapshot is $\simeq16$~mJy PSF$^{-1}$. As such, given the spatial oversampling, we suggest that we achieve near-thermal noise in our $P(\phi)$ cubes.

Our noise maps indicate spatial noise variation, in two respects. Firstly, the noise rises toward the edge of each field, due to the MWA primary beam response. Secondly, the noise was found to be higher at the location of particularly bright sources. This is likely due to some combination of both un-deconvolved Faraday-space sidelobes as well as instrumental leakage contamination away from zero FD. Hence we needed to account for this position dependence to properly characterise sources in our field.

\section{Source Identification and Verification}
We rejected source-finding directly on the polarisation data for a number of reasons, including the non-Gaussian nature of polarisation image noise and a lack of 3D source-finding algorithms in the literature \citep[these reasons are discussed in detail by][]{Farnes2018}. 

Instead, we opted to search for linearly-polarised emission at the location of known Stokes I sources. We used the GLEAM Extragalactic Catalogue \citep{HurleyWalker2017} for reference, and followed similar methodology to \cite{VanEck2018} to identify and verify sources. In this section, we will discuss our procedure.

\begin{figure*}
\begin{center}
\includegraphics[width=0.9\textwidth]{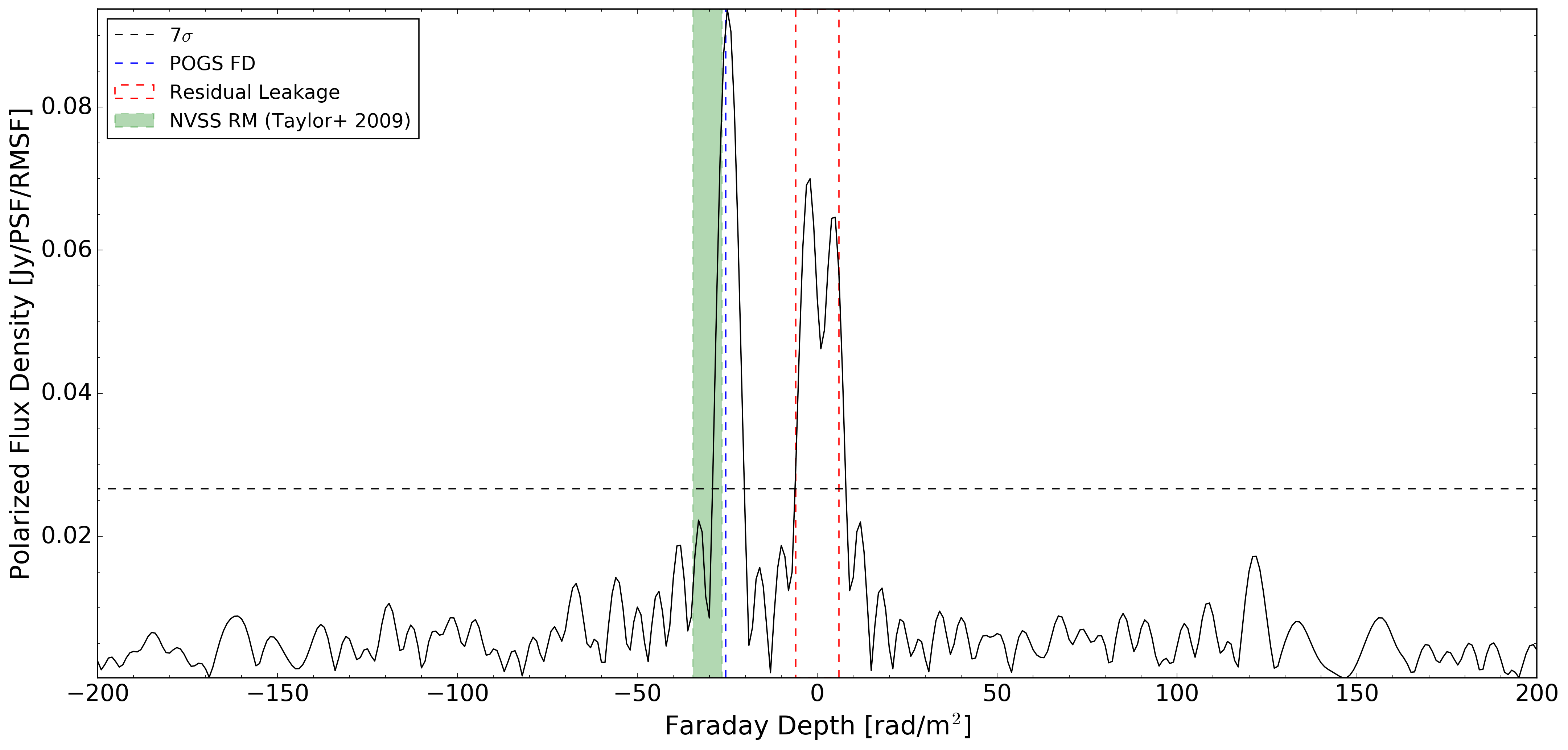}
\caption{Example Faraday depth spectrum of a real polarised source (GLEAM~J130025$-$231806). The measured POGS FD, NVSS RM \citep{Taylor2009} and leakage `zone of exclusion' are identified, as is the $7\sigma$ peak identification threshold. Note that the width of the NVSS RM box denotes the uncertainty on the measurement; the uncertainty on the POGS FD is not visible on this scale.}\label{fig:fdspectrum}
\end{center}
\end{figure*}

\subsection{Identifying candidate sources}
For each mosaic, we produced a catalogue containing a sub-sample of sources from the GLEAM Extragalactic Catalogue, containing sources with a 204~MHz integrated flux density in excess of 90~mJy located where the primary beam response greater than 30 per cent. This flux density cutoff was chosen in a compromise between attempting to minimise the number of spurious candidates and explore the polarimetric properties of the faint GLEAM source population. This yielded a catalogue of approximately 50,000 GLEAM sources in our survey region, at a typical surface density of around 8 deg$^{-2}$.

Following \cite{VanEck2018}, for each source we extracted a cube that sampled the entire Faraday depth range and covered a spatial region $6\sigma_{\rm{maj}}$\footnote{Where $\sigma_{\rm{maj}}$ is the major axis of an ellipse defined by the $1\sigma$ level of the image-plane PSF}. Pixels within the source FWHM were then identified as `on-source'. The source Faraday spectrum was then determined using the maximum polarised intensity of on-source pixels, as a function of FD. To measure the off-source spectrum, we overlaid the PSF on the location of each source, and determined the maximum polarised intensity of all pixels below the 1 per cent level, as a function of FD.

The source spectrum was then searched for peaks by identifying local maxima. For each peak identified, a small number of initial tests were performed to filter out spurious detections. Peaks were only identified as candidates if they fulfilled three criteria. Firstly, the peak flux density must be in excess of $7{\sigma_{\rm{P}}}$ (where ${\sigma_{\rm{P}}}$ is the local fitted noise in the FD cube, as discussed in Section~\ref{sec:noise}). Secondly, the peak flux density must be in excess of the off-source flux density plus $2{\sigma_{\rm{P}}}$. Thirdly, the Faraday depth at which the peak appears must be outside the instrumental leakage region\footnote{This was determined empirically by inspecting the FD spectra of both bright and faint Stokes I sources. In practice, this excluded peaks at $|\phi| \lesssim 6$~rad~m$^{-2}$}. 

It is likely that we are excluding real polarised sources with low FD by enforcing this third criterion. However, given our current ability to mitigate instrumental leakage, this is a necessary step. See Section~\ref{sec:nvssrm}. We present the FD spectrum for an example source that appears in our catalogue (GLEAM~J130025$-$231806) in Figure~\ref{fig:fdspectrum}. We note that there is some residual frequency-independent leakage (at the level of $\lesssim 1\%$), as well as some frequency-dependent leakage term (visible in the $|\phi| \lesssim 6$ rad m$^{-2}$ region). However, both leakage residuals are less significant than the real polarised emission from this source.

\subsection{Verifying candidate sources}\label{sec:fitting}
Our source-finding routine identified approximately 700 candidates. For each candidate, we extracted a region\footnote{Spanning the full FD range from the FD cube and six times the PSF major axis in both RA and Dec.} from the RM cube around the peak. We then fitted a 3D Gaussian (RA/DEC/FD) to this sub-cube.

Following \cite{VanEck2018} this Gaussian function was chosen to match the expected form (in both image- and FD-space) of an unresolved source, with background noise added in quadrature. This yielded a nine-parameter model:
\begin{itemize}
    \item Peak polarised intensity $(P)$
    \item Background polarised intensity $(C)$
    \item Image-plane centroids in pixel coordinates $(X,Y)$
    \item Image-plane semi-major $(\sigma_{\rm{maj}})$ and semi-minor $(\sigma_{\rm{min}})$ axes, measured as Gaussian $\sigma$
    \item Image-plane position angle $(PA)$
    \item Faraday depth centroid $(\phi)$
    \item Faraday depth width $(\sigma_{\phi})$, measured as Gaussian $\sigma$    
\end{itemize}

This model was optimised using the \texttt{scipy} `curve-fit' algorithm, employing a Levenberg-Marquardt solver. Initial guesses for each parameter were those measured during initial source identification. A small number of sources were identified with multiple peaks in FD; however, none of these were separated by a sufficiently narrow FD range that the peaks became blended (see Section~\ref{sec:results}). Our candidate list contained a large number of spurious candidates that (from inspection) were identifiable as sidelobes of the RMSF. The vast majority of these spurious candidates either failed to fit, or were poorly-constrained, and were subsequently eliminated. 

\subsection{Error quantification}
Following \cite{VanEck2018} we established a Monte-Carlo simulation to quantify the errors on our fitted model. This was done to account for the correlated noise in our cubes (which is not the accounted for by `curve-fit') caused by the limited resolution of our PSF and RMSF.

The full method is discussed in detail by \cite{VanEck2018}. Briefly, however, we performed a FFT of our 3D PSF+RMSF. Each of the real and imaginary components of this FFT was then multiplied by a random complex number drawn from a separate Gaussian distribution. These were then Fourier transformed into a FD cube, with the imaginary component discarded, and the standard deviation scaled to match that of our real data. 

For each candidate, we established 1000 realisations of noise, adding the best-fit source model to each, and performed our fitting routine on each realisation. We then used the standard deviation of the fit results to estimate the measurement uncertainty. Whilst these Monte-Carlo errors may not perfectly capture the true uncertainties \citep[see][]{VanEck2018} they proved to be a powerful tool with which to eliminate false positives from our catalogue, and they broadly match the scatter observed in our both our data and that of \cite{VanEck2018}. 

The theoretical uncertainty on the measured FD of a source is inversely proportional to the SNR of the detection \citep[e.g.][]{Brentjens2005} according to:
\begin{equation}\label{eq:drm}
    \delta{\phi} \propto \frac{ \Delta \phi }{ 2 \times {\rm{SNR}} }
\end{equation}
We note that our MC uncertainties are typically comparable to the theoretical uncertainty for sources with high SNR. For sources with lower SNR $(\lesssim 20)$ the MC uncertainties are typically 30 per cent larger than predicted by Equation~\ref{eq:drm}.

\subsection{Candidate evaluation}
From Equation~\ref{eq:rmparms3}, we know that our POGS sample will only be sensitive to very Faraday-thin sources, as the maximum scale we can recover in FD is $1.9$ rad m$^{-2}$. Note that this maximum scale size is $\ll \Delta\phi$. As such, any sources that appear to be significantly more extended in FD are likely to be spurious and should be excluded. We rejected any candidates more than $1.9\times$ the extent of the RMSF.

Our FD cubes possess modest resolution and sensitivity (approx. 3 arcminutes and $\sim3-5$ mJy PSF$^{-1}$ RMSF$^{-1}$, respectively). While our use of an inner $uv$-cut has reduced contamination from diffuse foreground, visual inspection revealed a number of Stokes I sources appear to be co-located with patches of diffuse Galactic polarised emission.

As such, we have assumed that candidates with a fitted major axis significantly larger than the PSF are the result of such diffuse foregrounds. Candidates with fitted size more than $1.8\times$ the PSF size were rejected. This threshold was empirically determined by visually examining both promising candidates and suspected spurious sources.

However, we found the most effective parameter with which to discriminate between real and spurious candidates was the position error, defined as the quadrature sum of the fitted X- and Y-centroid uncertainties. For sources which were identified by-eye, the position error was typically $\lesssim 0.08$ pixels. However, for sources which were visually identified as spurious, the position error was $\gtrsim 0.2$ pixels. We selected an upper-limit acceptance threshold of 0.12 pixels for the position uncertainty.

Note that we did not require polarised sources to lie coincident with the peak in Stokes I. Given the moderate resolution of the MWA, many Stokes I sources that appear compact at 3~arcminute resolution become resolved into multiple sources at higher resolution. As such, a real polarised source in our catalogue may be associated with a single component in total intensity that is unresolved in GLEAM. Indeed, from inspection, this was found to be true -- see Section~\ref{sec:morphology}, or the postage stamp images presented later in this paper.

For a fraction of candidate sources, initial peaks were identified that lay close to the leakage exclusion region. As a precaution, we inspected the FD spectra for these sources. The fitted FD peaks for a small sub-sample had shifted to within the exclusion region; such sources were removed from our catalogue.\footnote{The fact that these fitted peaks shifted compared to the initial identification is a strong indicator that these are indeed spurious candidates.}

We note that an improved frequency-dependent leakage routine may reduce the residual low-FD leakage seen in the Faraday spectra for some sources (as seen in Figure~\ref{fig:fdspectrum}) as well as further improving the zero-FD leakage. Some previous studies \citep[e.g.][]{Lenc2017,OSullivan2018} have mitigated leakage by using bright sources to perform in-field calibration; however, this was not practical for this large-scale reprocessing. Image-plane leakage corrections have been employed at low frequencies by \cite{Lynch2017} and in the circular polarisation survey performed by \cite{Lenc2018}. To our knowledge, this is the first large-scale application of such techniques in linear polarisation at low frequencies, though we note that \cite{Condon1998} use holography observations of strong sources to derive an empirical image-plane leakage correction for the NVSS.

\subsection{Final measurements}
All candidates that conformed to the criteria defined in the previous section, we considered real. These are henceforth referred to as `sources'. For each source, the final FD was determined using the mean of all detections. Following the assumption that the measured background is largely dominated by noise, we determined the final polarised flux density measurements through the quadrature subtraction of the noise from the fitted flux density \citep[e.g.][]{George2012}.

\section{Results}\label{sec:results}
We present our catalogue of linearly-polarised sources in Table~\ref{tab:catalogue}. Our catalogue contains 81 sources above a limiting polarised flux density of 18~mJy PSF$^{-1}$. Additionally, we find 30 candidates that cannot yet be conclusively discriminated from foreground emission; these will be discussed in a follow-up paper when we consider the full POGS catalogue. Note that in Table~\ref{tab:catalogue} the quoted Right Ascension and Declination are the location of the polarised peak, rather than the Stokes I centroid from the GLEAM catalogue, as some sources exhibit an offset between the peaks of total intensity and polarised intensity.

Cross-examination with the NVSS reveals that these sources are often compact-double radio sources, with the polarised emission typically clearly associated with a single NVSS continuum source. For these double sources, we used Aegean 2.0 \citep{Hancock2012,Hancock2018} to re-fit the GLEAM `white' mosaics (mosaicked and stacked GLEAM images covering the frequency range 204--232~MHz) using the NVSS catalogue entries as a prior. We have indicated in Table~\ref{tab:catalogue} where this was the case.

\begin{figure*}
\begin{center}
\includegraphics[width=\textwidth]{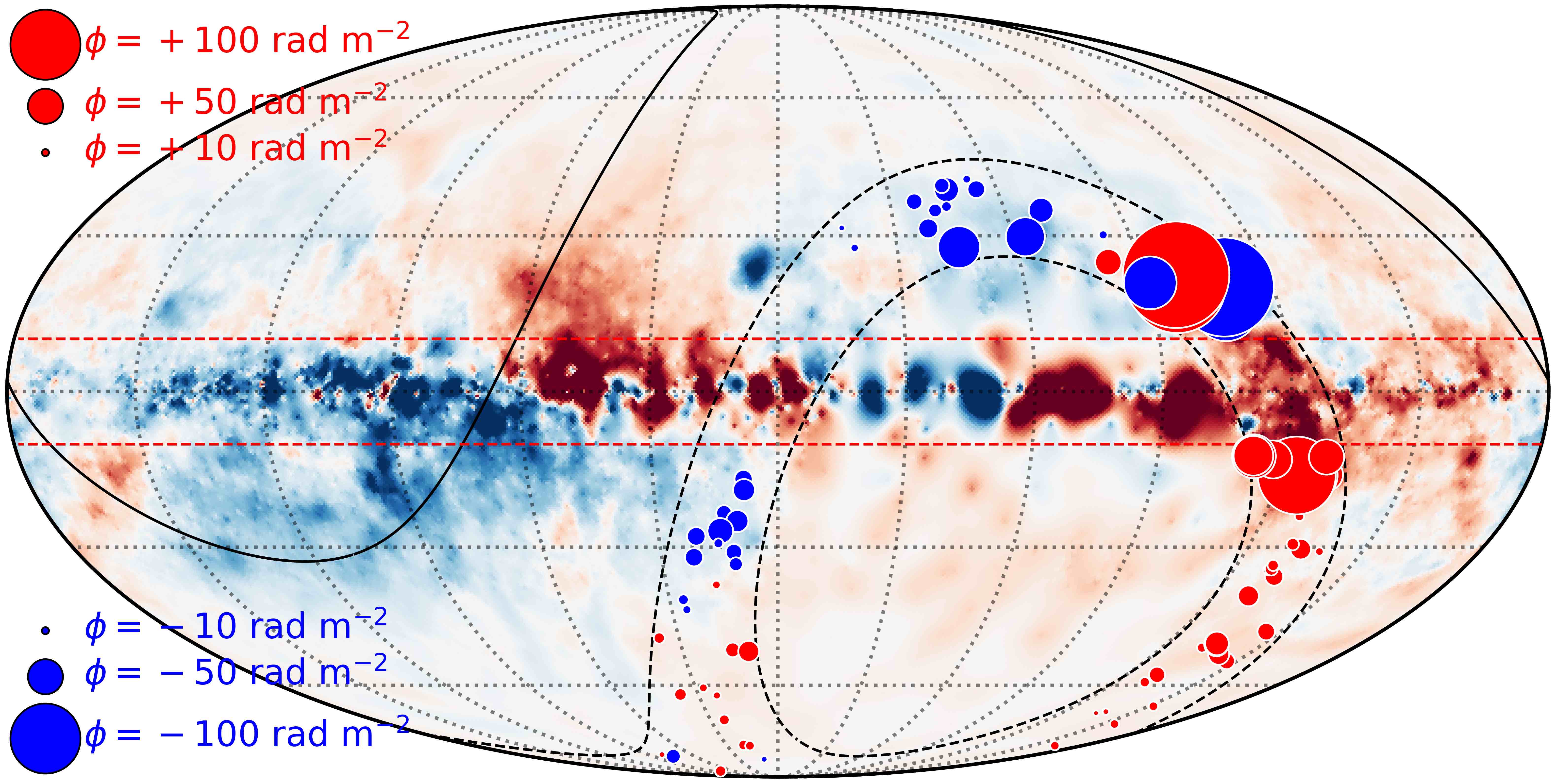}
\caption{Sky surface distribution of sources in our POGS catalogue. Background colourscale is the Galactic FD map from \cite{Oppermann2012} saturating at $|\phi|=200$~rad~m$^{-2}$. Markers denote the location of POGS sources, with the colour indicating the sign of the FD (blue is negative FD, red is positive). The markers are scaled according to the magnitude of the FD, with some example marker sizes shown in the inset. The black dashed and solid lines denote the limits of the region covered in this paper, and the limits of the GLEAM survey area, respectively. The red dashed lines denote the region excluded from the GLEAM Extragalactic Catalogue \citep[$|b| < 10\degree$;][]{HurleyWalker2017}.}\label{fig:skydist}
\end{center}
\end{figure*}

A small number of sources in our catalogue exhibited evidence of multiple peaks in FD; however, at most one of these peaks passed all tests. Two polarised sources were identified in connection with multiple GLEAM sources - in each case, the incorrect association was removed from our catalogue. Eleven sources in our catalogue are associated with seven large-angular-scale (LAS) or giant radio galaxies (GRG). We discuss four of these sources individually (with postage stamp images) later in this paper; we present postage stamp images of the remaining sources at the end of this manuscript in Figure~\ref{fig:multi_postage}. Given that we are essentially only sensitive to compact, Faraday-thin sources in this work, all postage stamp images are slices through our FD cubes at the peak FD, rather than a peak polarisation or moment map. This may become more important with the extended MWA configuration \citep{Wayth2018} with the improved sensitivity and resolution we expect to achieve.

\section{DISCUSSION}
\subsection{Source sky density}
Whilst we cannot yet quantify our completeness -- largely as a result of our source-finding methodology, practice of excluding leakage-like FD, as well as various depolarisation mechanisms -- we can estimate a lower limit to the surface density of sources on the sky at this frequency.

Our sample of the GLEAM survey covers approximately 6400 square degrees. In this region, the GLEAM Extragalactic Catalogue \citep{HurleyWalker2017} contains approximately 90,800 sources. With 81 sources in our catalogue, we find the surface density of polarised sources to be $0.013~{\rm{deg}}^{-2}$, or one source per 79 deg$^2$, above a flux density threshold of 18~mJy PSF$^{-1}$. Assuming this is representative of the entire sky, we expect to detect of the order of 300 sources across the region covered by the GLEAM Extragalactic Catalogue. We note that this is slightly higher surface density than that measured by \cite{Lenc2016}, where comparable sensitivity was achieved in linear polarisation at 154~MHz.

This limiting flux density is intermediate to previous low-frequency polarimetric work with LOFAR \citep[$0.5-1$~mJy; e.g.][]{Mulcahy2014,VanEck2018} and the MWA \citep[350~mJy;][]{Bernardi2013}. The catalogue of \cite{VanEck2018} covers a region of 570~deg$^2$, containing six sources with flux densities above our threshold\footnote{Where we have taken a typical synchrotron spectral index $\alpha = -0.8$ to scale our 216~MHz minimum flux density to 150~MHz.}. Our surface density suggests we would recover seven sources in such an area, so we suggest that we find comparable source density to \cite{VanEck2018}. We note that we are at slightly higher frequency than \cite{VanEck2018} and as such we might expect a slightly higher true source density. However, we suggest that this comparison is limited by some combination of small-number statistics and cosmic variance. Note that our sample is less limited by cosmic variance as we cover a significantly larger region of sky. Additionally, like \cite{VanEck2018} we are certainly incomplete at low FD due to our practice of avoiding the `leakage region'.

We note an additional source of incompleteness in our survey, caused by the fact that the GLEAM catalogue only contains sources at $|b| \geq 10\degree$ \citep[see][]{HurleyWalker2017}. As such, some regions of sky that were imaged during this work were not searched for sources. We will revisit this once the GLEAM Galactic catalogue becomes available; as such our prediction of a total 300 sources may be a lower limit. Whilst many sources viewed through the Galactic plane would likely be depolarised at 3~arcmin resolution, we anticipate that we will detect additional pulsars and/or pulsar candidates from this region. Furthermore, an additional source of incompleteness arises when discriminating between chance alignment of diffuse Galactic emission and low SNR polarised sources that lie coincident with patches of Galactic emission. Whilst our imaging settings (described in Section~\ref{sec:processing}) were selected to mitigate Galactic contamination, some residual emission is still visible in our FD cubes. All sources that could not be conclusively separated from Galactic emission were excluded from this catalogue; however, they will be re-examined in Paper II (Riseley et al., in prep) where will present the all-sky catalogue.

Figure~\ref{fig:skydist} presents the sky surface distribution of sources in our catalogue, overlain on the Galactic FD map of \cite{Oppermann2012}. From Figure~\ref{fig:skydist}, we note two principal features. Firstly, the sign of the FD (indicated by the colour, with blue being negative FD and red being positive FD) measured for our sources are broadly consistent with the Galactic foreground, suggesting that the dominant contribution to the observed FD is caused by our own Galaxy. Based on this, we tentatively suggest that the non-uniform surface density exhibited in Figure~\ref{fig:skydist} is likely caused by a physical depolarisation mechanism, rather than our practice of excluding low-absolute-FD sources, as the overwhelming majority of the Galactic FD has $|\phi| > 6$~rad~m$^{-2}$. This will be investigated further in Paper II.

Secondly, we note an apparent clustering of sources with large absolute FD. This cluster is in close proximity to the Northern H$\alpha$ arc of the Gum Nebula \citep[e.g.][]{Stil2007,Oppermann2012,Purcell2015}. \cite{Vallee1983} identified a large magnetic bubble in this region, associated with the H$\alpha$ emission through analysis of the average RM in concentric rings centered on the Gum Nebula. Within approximately 20$\degree$ of the Gum Nebula, the average RM was significantly higher than that further out \citep[up to $|{\rm{RM}}| \simeq 200$ rad m$^{-2}$;][]{Vallee1983}. The overdensity of sources with high absolute FD seen in this region of Figure~\ref{fig:skydist} suggests that we may also detect this feature, although this overdensity lies at approx. $30\degree$ radius from the centre of the Gum Nebula.

\subsection{Comparison with NVSS rotation measures}\label{sec:nvssrm}
Based on the NVSS catalogue\footnote{Which contains measurements in both total intensity and linear polarisation \citep{Condon1998}.}, \cite{Taylor2009} derive RMs for 37,543 sources at Declinations higher than $-40\degree$. We have searched the entire sky between $-37\degree < \delta < -17\degree$ and $|b| > 10\degree$. In this region, there are approximately 4700 NVSS sources with RMs in the range we have searched in this initial study $(6 \lesssim|\phi|~[{\rm{rad~m}}^{-2}]\lesssim 200)$.

\begin{figure}
\begin{center}
\includegraphics[width=0.49\textwidth]{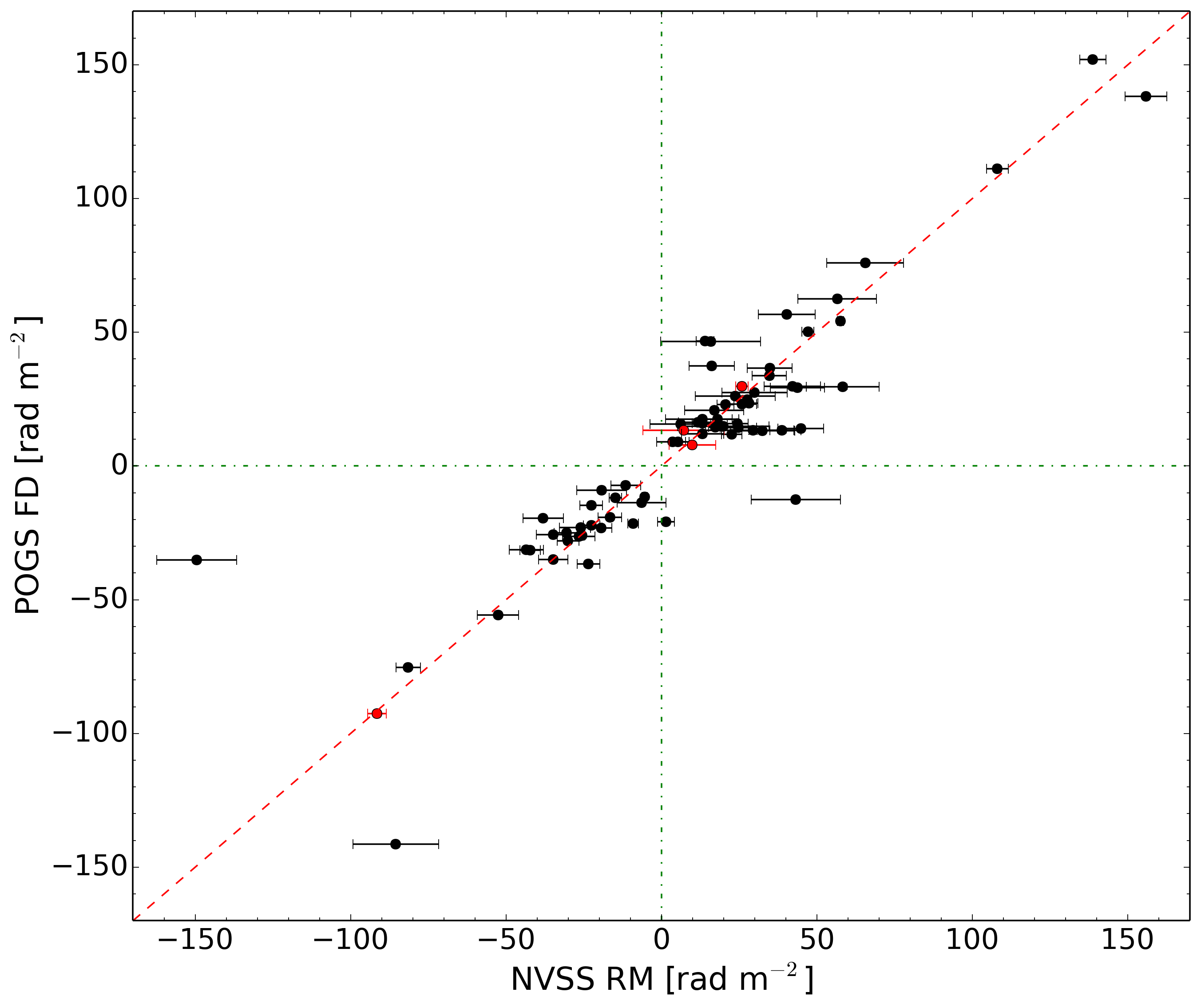}
\caption{Comparison of rotation measure (RM) and Faraday depth (FD) for sources common to both the \cite{Taylor2009} RM catalogue and our POGS catalogue, respectively. The dashed diagonal line denotes the 1:1 correspondence. Dot-dashed lines mark zero FD. Red points denote sources in Table~\ref{tab:catalogue} where the polarised component has no clear association with a single NVSS source.}\label{fig:rms}
\end{center}
\end{figure}

We note that approx. 17 per cent of NVSS polarised sources in the overlap region have RMs in the instrumental leakage `zone of exclusion'. As a result of this, our catalogue is certainly incomplete at low FD. Of the 81 sources in our POGS catalogue, 71 have counterparts in the \cite{Taylor2009} catalogue. We present the comparison between NVSS RM and POGS FD in Figure~\ref{fig:rms}, which shows a broad consistency between the FD we recover and the RM measured by \cite{Taylor2009}. 

However, for a small number of sources, the difference between NVSS RM and POGS FD is greater than can be explained by measurement uncertainties. There are a number of possible explanations for this. The $n\pi$ ambiguity in the NVSS RMs (as a result of having only two $\lambda^2$ samples) cannot account for these differences, as a single wrap between frequency bands would manifest as an RM shift of $\pm652.9$ rad m$^{-2}$ \citep{Taylor2009}. Given that we are essentially only sensitive to Faraday-thin sources, it is possible that a sub-sample of the population detected here may have some extended FD component that is being probed at 1.4~GHz but is depolarised at 200~MHz. Alternatively, variability may play some role.

An additional explanation for this may be differences in the spectral index structure of the polarised emitting regions of these sources. With one exception, all sources which exhibit this stark difference between POGS FD and NVSS RM are associated with AGN (see Table~\ref{tab:class}). It is well-established that AGN hotspots often occur as complexes of emission \citep[e.g.][]{Laing1982,Orienti2012} -- as such, a steep-spectrum polarised component detected by the MWA may have a different measured FD than that measured for a flat-spectrum component detected by the NVSS. All are unresolved by both the NVSS and MWA -- high-resolution observations would be required to confirm this scenario. The sole exception to this is the pulsar PSR~B0628$-$28 (see Section~\ref{sec:psr}) for which our measured FD is in agreement with the measurement from the ATNF Pulsar Catalogue \citep{Manchester2005}.

\subsection{Depolarisation analysis}

\subsubsection{Polarisation fraction}
Investigating the variation of polarisation fraction $(m)$ with frequency can allow us to investigate the FD structure inside polarised sources \citep[e.g.][]{Farnes2014} via depolarisation. This in turn can allow us to infer information about the magnetic field structure inside radio sources \citep[e.g.][]{Anderson2016,OSullivan2018}. Additionally, depolarisation can tell us about the properties of the magnetoionic medium along the LOS, both of the Milky Way as well as intervening systems \citep[e.g.][]{Haverkorn2008,Anderson2015}.

\begin{figure}[!t]
\begin{center}
\includegraphics[width=0.48\textwidth]{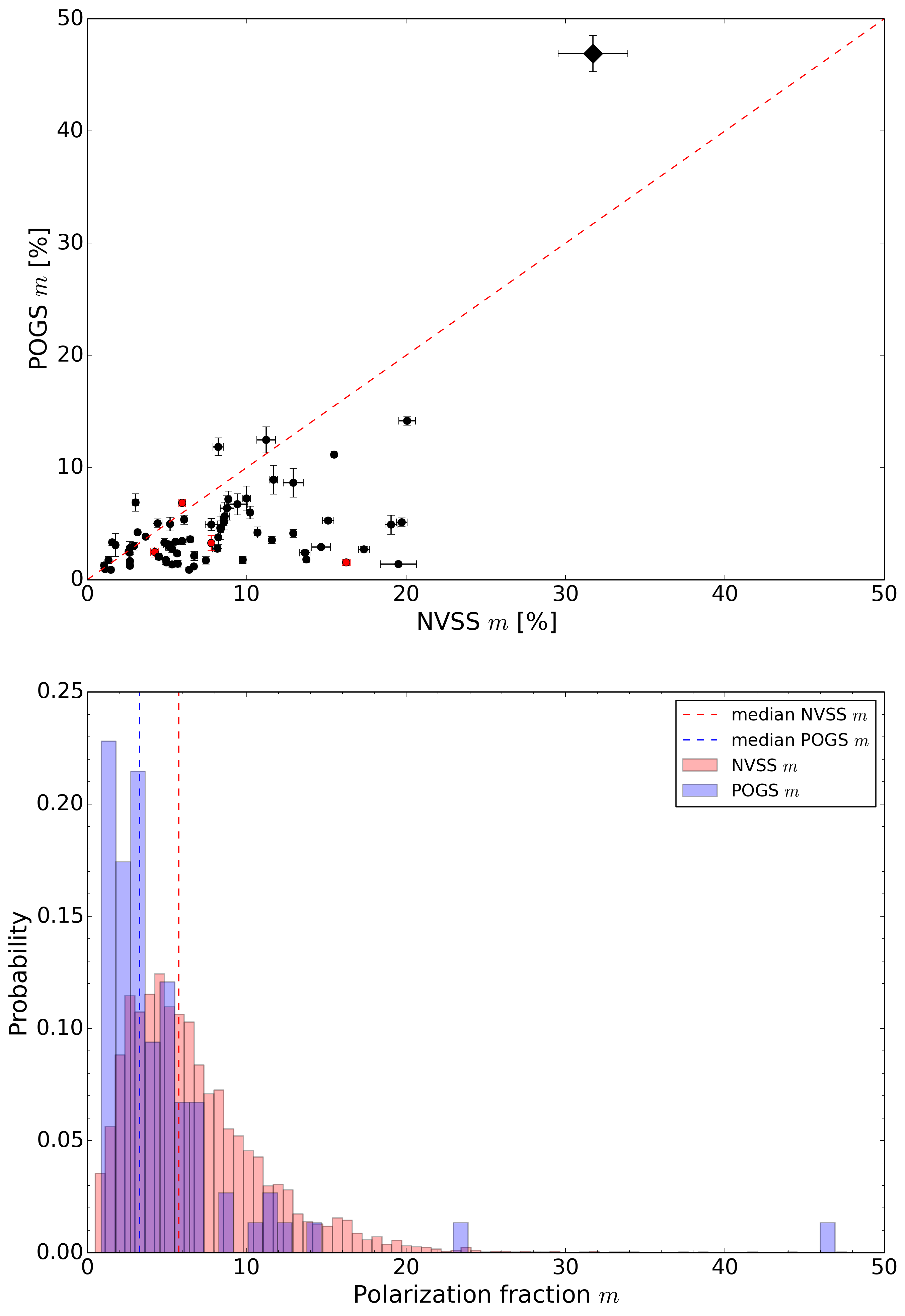}
\caption{Fractional polarisation $(m)$ distribution for our POGS catalogue and the \cite{Taylor2009} catalogue. \emph{Top panel:} depolarisation for sources common to both catalogues. The dashed line denotes the 1:1 correspondence. Red points denote sources in Table~\ref{tab:catalogue} where the polarised component has no clear association with a single NVSS source. The diamond denotes the position of the pulsar PSR~B0628$-$28. \emph{Bottom panel:} histogram of $m$ for all sources in the POGS catalogue (blue) and for the \cite{Taylor2009} catalogue in the same region of sky and FD range (red). The dashed lines denote the median polarisation fraction at each frequency: 3.29\% at 216~MHz and 5.74\% at 1.4~GHz.}\label{fig:polfrac}
\end{center}
\end{figure}

We present the fractional polarisation $m$ for sources common to both our POGS catalogue and the NVSS RM catalogue in Figure~\ref{fig:polfrac} (top panel). From Figure~\ref{fig:polfrac} it is clear that the majority of sources exhibit reduced polarisation fraction at 200~MHz compared to 1.4~GHz. However, 14 sources in our catalogue are consistent with a depolarisation ratio $m_{\rm{216MHz}}/m_{\rm{1.4GHz}} > 1$. This is often known in the literature as either `repolarisation' or `anomalous depolarisation'; in this work we will adopt the former term. Repolarisation has been observed in both spiral galaxies \citep[e.g.][]{Horellou1992} and AGN jets \citep[e.g.][]{Sokoloff1998}.

One source that appears to exhibit repolarisation is the pulsar PSR~B0628$-$28. We discuss this source further in Section~\ref{sec:psr}. Furthermore, the hotspots of one of the LAS radio sources in our catalogue also exhibit repolarisation; we discuss this source later in Section~\ref{sec:pks0707}. Furthermore, we note that while GLEAM~J023512$-$293622 (ID 01-08) appears to exhibit repolarisation, the NVSS polarisation image of this source reveals a single polarised component associated with the dominant total intensity component. It is possible that the \cite{Taylor2009} entry for this source is spurious, as their fitting routine resulted in polarised emission being associated with both total intensity components. 

Of the remaining sources, seven exhibit `compact-single' morphology, whereas two exhibit `compact-double' morphology (see Section~\ref{sec:morphology} for details). While the fact that these compact-double sources exhibit repolarisation could indicate an incorrect association, we consider this unlikely, as these sources\footnote{IDs 01-25 and 01-40 in Figure~\ref{fig:multi_postage} and Tables~\ref{tab:catalogue} and \ref{tab:class}.} are clearly associated with a single NVSS component. 

Instead, for one of these sources (01-40, GLEAM~J092317$-$213744) we note that the NVSS RM $(-85.5\pm13.3~{\rm{rad~m}}^{-2})$ is also highly discrepant with our measured FD $(-141.3\pm0.3~{\rm{rad~m}}^{-2})$. Taking our measured FD as true for this source, Figure~1 of \cite{Taylor2009} suggests that the polarised flux measured at 1.4~GHz will be underestimated by approximately $20-25\%$. This would put the $m_{\rm{1.4GHz}}$ closer to 2.1\%, although our measured $m_{\rm{216MHz}}=3.3\%$ would still be repolarised.

Two principal explanations exist for repolarisation: either helical magnetic field structure \citep[e.g][]{Urbanik1997,Sokoloff1998} or the presence of multiple unresolved components within a given synthesised beam \citep[e.g.][]{Farnes2014b}. In the latter scenario, the different components may have different polarisation fractions and different spectral indices in order to present repolarisation behaviour. Given the moderate resolution of the MWA, we suggest that perhaps the latter scenario is the more likely of the two for these repolarising sources. Approximately half of the repolarising sources also exhibit FD differences that are too large to be explained by measurement uncertainties in the NVSS RMs, which provides some evidence in support of this interpretation. However, we also note that Faraday-space interference effects from multiple components would likely only cause repolarisation over a limited frequency range -- outside which the source would visibly depolarise \citep[e.g.][]{Slysh1965,OSullivan2012}. Further broad-band observations of these sources at higher frequencies would be required to investigate further.

The lower panel of Figure~\ref{fig:polfrac} also presents a histogram of the polarisation fraction for all sources in the POGS catalogue and those in the NVSS RM catalogue that i) are in the region considered in this work and ii) have measured RMs in the range $6<|{\rm{RM}}|<200$ rad m$^{-2}$. From Figure~\ref{fig:polfrac}, the polarisation fraction distribution of our catalogue is dominated by sources with low $m$. We note that the NVSS RM catalogue is likely incomplete at very low $m$, as \cite{Taylor2009} excluded sources with $m < 0.5\%$, which may explain some of the fall-off seen in Figure~\ref{fig:polfrac}. 

For NVSS sources in the parameter space explored in this work, the median 1.4~GHz polarisation fraction is $\tilde{m} = 5.7\%$. Note that this is comparable to the typical 1.4~GHz polarisation fraction for reliably-detected polarised sources in the NVSS catalogue, $\tilde{m}_{\rm{1.4~GHz}} \simeq 8.1\%$ \citep{Condon1998}. The median POGS source has a polarisation fraction $\tilde{m}_{\rm{216~MHz}} = 3.3\%$ and $\tilde{m}_{\rm{1.4~GHz}} = 6.7\%$, for a depolarisation ratio $\tilde{m}_{\rm{216~MHz}}/\tilde{m}_{\rm{1.4~GHz}} \simeq 0.5$. This provides further evidence that the population of sources detected in polarisation by the MWA are hotspots of radio galaxy lobes: for example, the median polarisation fraction of all sources identified by \cite{Hammond2012} as active galaxies (where the radio emission is lobe-dominated) is $\tilde{m} = 7.8\%$.

\subsubsection{Polarisation fraction vs. Faraday depth}
\begin{figure}[!t]
\begin{center}
\includegraphics[width=0.48\textwidth]{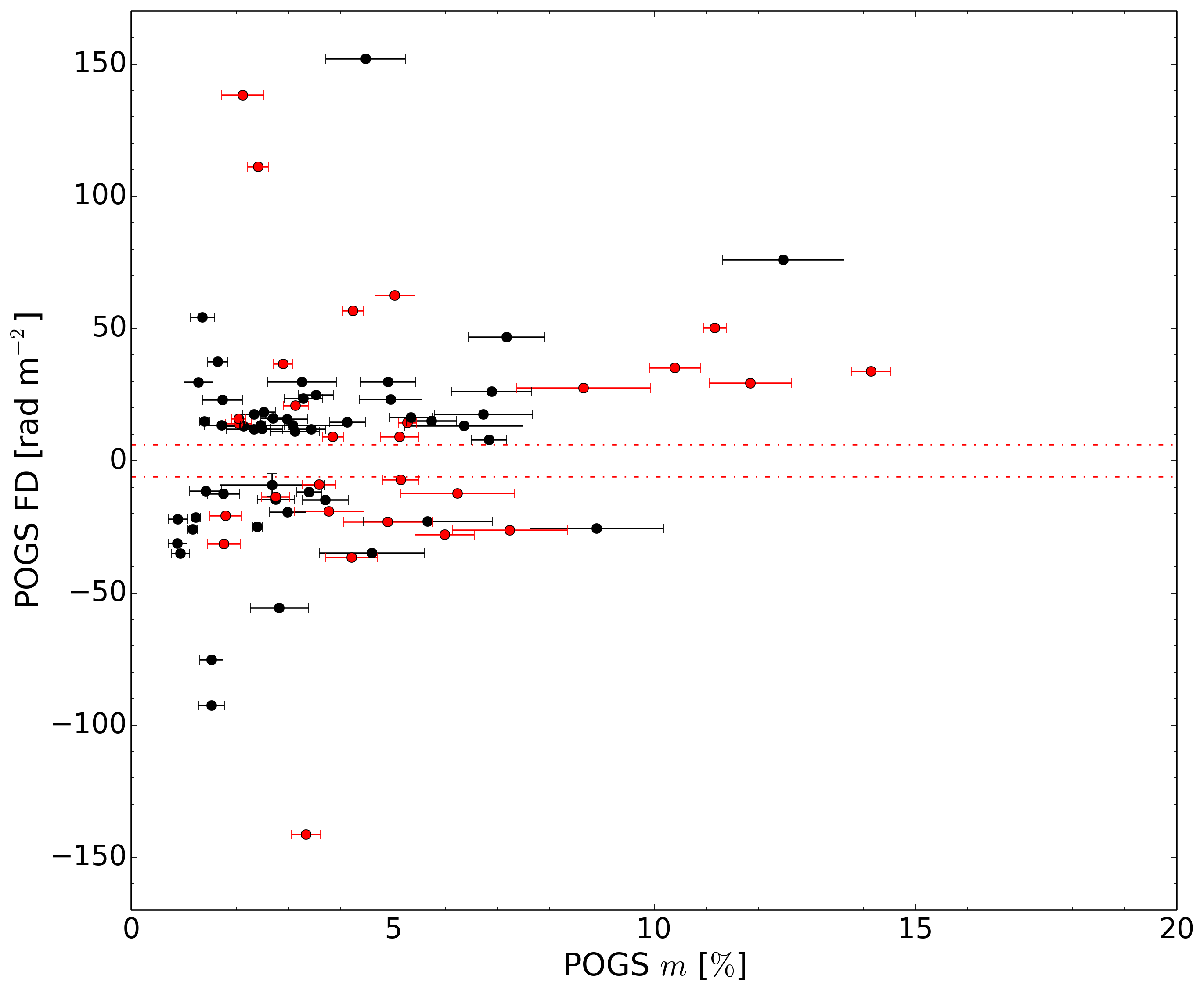}
\caption{Faraday depth as a function of fractional polarisation, $m$ for sources in our catalogue. Red points denote sources for which the Stokes I flux density was measured using priorised fitting with Aegean 2.0 \citep{Hancock2018}; black points denote sources where the GLEAM 212~MHz Stokes I flux density measurement was used. Dot-dashed red lines denote the instrumental leakage `zone-of-exclusion'. Note that we have excluded the known pulsar and pulsar candidate from this plot -- see Section~\ref{sec:psr}.}\label{fig:mfd}
\end{center}
\end{figure}

Following \cite{Hammond2012} we also investigated the dependence of our measured FD on the measured polarisation fraction for sources in our catalogue. We present this dependence in Figure~\ref{fig:mfd}.

Whilst we are slightly limited by small number statistics, from Figure~\ref{fig:mfd}, it appears that we observe a broadly similar trend to \cite{Hammond2012} in their sample of NVSS RMs. Namely, Figure~\ref{fig:mfd} shows that the population of sources with lower polarisation fraction tend to exhibit broader standard deviation in their measured FD; conversely, sources with greater polarisation fraction tend to exhibit less standard deviation in FD. 

As suggested by \citeauthor{Hammond2012}, this behaviour is likely astrophysical rather than instrumental, associated with a depolarisation mechanism. Given the channel bandwidth used in this work, we are not limited by bandwidth depolarisation, as $|\phi_{\rm{max.}}| \simeq 1900$~rad~m$^{-2}$. Faraday depth depolarisation typically occurs where the emitting region is mixed with a Faraday-rotating medium (also known as a `Burn slab') causing complex and/or extended FD structures \citep[e.g.][]{Hammond2012,OSullivan2013,Kaczmarek2018}. However, given that the MWA is only sensitive to sources that are essentially Faraday-thin (from Equation~\ref{eq:rmparms3}, the maximum scale in FD we can recover is around 1.9~rad~m$^{-2}$) any `Burn slab' sources will very rapidly become undetectable.

Additionally, as shown by \cite{Hammond2012} there is little evolution in FD with redshift, which suggests that the observed FD signal is caused by a foreground. Indeed, previous works have shown that the Galactic foreground FD varies significantly on small scales, leading to higher absolute FD values and increased depolarisation even for unresolved sources \citep[e.g.][]{Haverkorn2008}. As such, given our moderate resolution of 3~arcminutes, we suggest that beam depolarisation (whereby the FD varies on scales smaller than our PSF) is the most likely explanation for this trend. We will revisit this in Paper II, where we our larger all-sky sample will provide the necessary statistics for a more quantitative analysis.

\section{Source classification}

\subsection{Morphological classification}\label{sec:morphology}
Broadly-speaking, we can classify the sources in our catalogue based on their radio morphologies from both the GLEAM and NVSS data. We divide our sources into three categories: compact-single (CS), compact-double (CD) and extended-double (ED). 36 sources in our catalogue exhibit CS morphology: unresolved or only partially-resolved at the resolution of both GLEAM and the NVSS. For these sources, the polarised peak generally coincides with the Stokes I centroid. 

Similarly, 34 sources exhibit CD morphology. These are typically unresolved by GLEAM, but show dumbbell morphology at the NVSS resolution. With only two exceptions, all CD sources show an offset between the GLEAM Stokes I centroid and the polarised peak -- which lies coincident with one of the NVSS components. There are 11 ED sources in our catalogue. These correspond to emission from different regions of seven LAS radio galaxies. Four of these LAS radio galaxies (PMN~J0351$-$2744, PKS~J0636$-$2036, PKS~0707$-$35, and ESO~422$-$G028) have two polarised components, and we discuss these further in Section~\ref{src:lasrg}. The remaining three have only a single polarised component, associated with a hotspot, and will be considered further as part of the full POGS sample in another paper. These sources we identify in Table~\ref{tab:catalogue}, Table~\ref{tab:class} and Figure~\ref{fig:multi_postage} as 01-03, 01-04 and 01-79.

\subsection{Host properties}
The majority of sources in our catalogue have candidate hosts identified from previous surveys. We present the cross-identifications in Table~\ref{tab:class}, but we summarise our classifications below:

\begin{itemize}
	\item AGN : 51 sources 
	\item Pulsar : 1
	\item Unknown : 29
\end{itemize}

Note that under the `AGN' bracket, we include objects with a clearly-classified optical host (typically Seyfert II or LINER galaxies) as well as those without a more well-defined classification. Eleven `ED' sources are hosted by seven AGN\footnote{Four of which are discussed in Section~\ref{src:lasrg}} -- that is to say, there are four extended radio sources associated with AGN where both hotspots are detected in polarisation, and three AGN with only a single polarised hotspot -- the remaining AGN host radio emission that is approximately evenly divided into `CD' and `CS' morphology. Of the 29 sources without an identified host, 16 exhibit CD morphology; these are likely AGN-type objects. All have GLEAM in-band spectra $\alpha^{\rm{232~MHz}}_{\rm{76~MHz}} \lesssim -0.7$.

\begin{figure*}
\begin{center}
\includegraphics[width=0.8\textwidth]{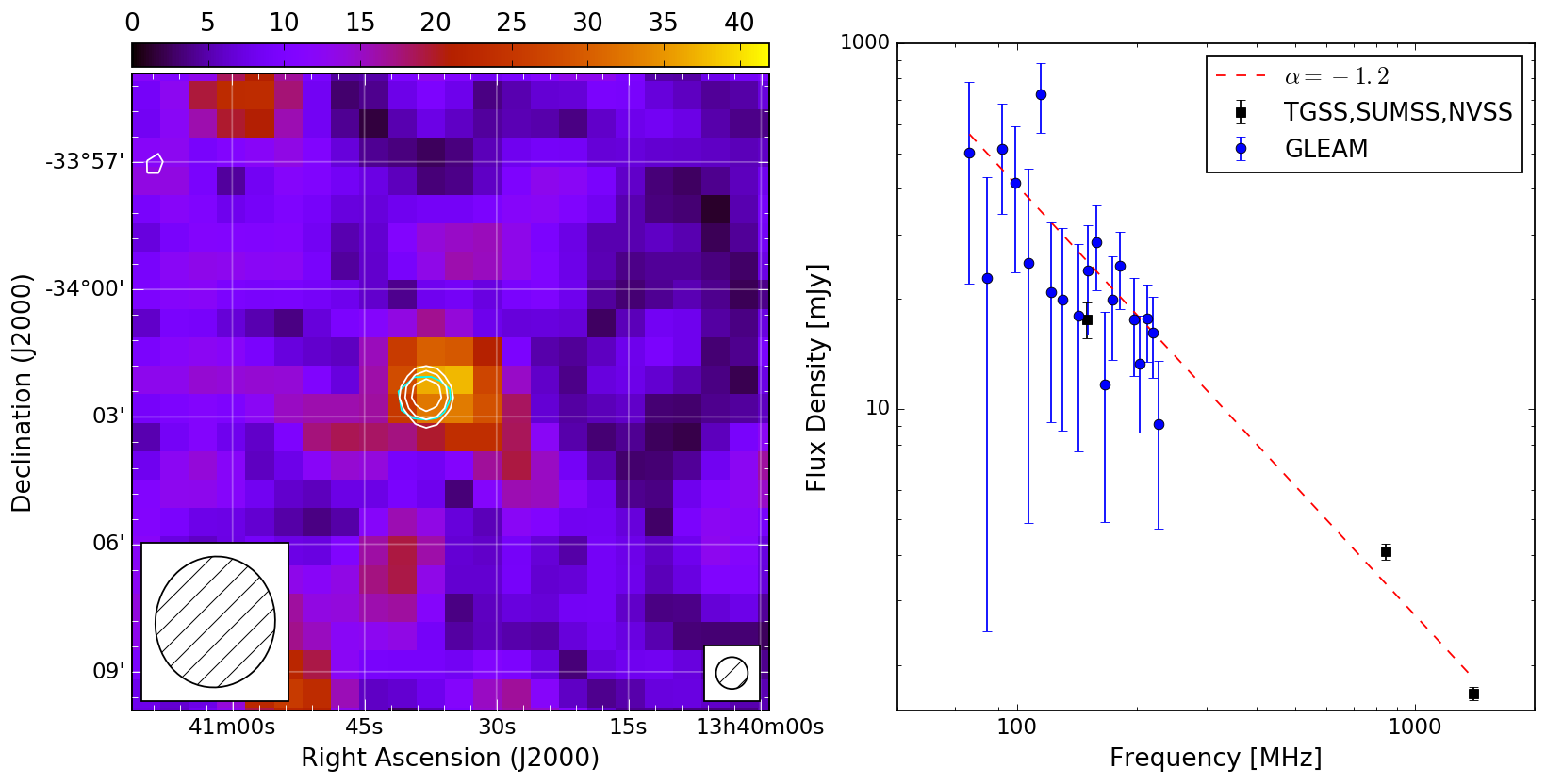} 
\caption{Postage stamp image (left) and spectral index (right) for our pulsar candidate, GLEAM~J134038$-$340234 (01-54 in Tables~\ref{tab:catalogue} and \ref{tab:class}). \emph{Left:} Colour scale is linearly-polarised intensity (in mJy PSF$^{-1}$) cut from a slice through our cube at $\phi=-60$~rad~m$^{-2}$. Cyan contours are GLEAM continuum from the "white" mosaic (204--232~MHz) starting at $5\sigma_{\rm{local}}$ and scaling by a factor two, where $\sigma_{\rm{local}}$ is the `wide-band' rms \citep[see][]{HurleyWalker2017}. White contours are NVSS Stokes I, starting at 2.25~mJy PSF$^{-1}$ $(5\sigma)$ and scaling by a factor two. The relative beam sizes of GLEAM and the NVSS are indicated by the hatched ellipses. \emph{Right:} blue (black) points are measured flux densities from GLEAM (the literature) respectively, as indicated in the inset. Dashed red line is the best-fit power-law spectral index $\alpha = -1.2$.}
\label{fig:psrcand}
\end{center}
\end{figure*}

\subsection{Pulsars}\label{sec:psr}
\subsubsection{GLEAM J063049$-$283438}
This source is the well-known pulsar PSR~B0628$-$28. Multi-frequency polarimetric observations suggest the polarisation fraction decreases steadily between 240~MHz and 3.1~GHz \citep{Johnston2008}. This trend is also consistent with much of the broader pulsar population \citep{Noutsos2015}. This is believed to be the result of the superposition of two orthogonal propagation modes \citep{Manchester1975}. We refer the reader to \cite{Noutsos2015} for further discussion of the physics behind the broad-band polarisation properties of pulsars. Previous observations with the MWA at 154~MHz found evidence of scintillation \citep{Bell2016}. However, this would not typically affect the fractional polarisation. 

From Figure~\ref{fig:polfrac}, our results suggest significantly higher fractional polarisation at 216~MHz $(m=46.9\pm1.6\%)$ compared to 1.4~GHz $(m=31.72\pm2.19\%)$, consistent with the results of \cite{Johnston2008}. This is consistent with the prediction that pulsars should be almost entirely Faraday-thin, although a larger sample would be required to test this further. We note that the NVSS RM $(+15.8\pm16.1~{\rm{rad~m}}^{-2})$ is highly discrepant with our measured FD $(+46.6\pm0.1~{\rm{rad~m}}^{-2})$, which is consistent with the RM from the ATNF Pulsar Catalogue\footnote{\url{https://www.atnf.csiro.au/research/pulsar/psrcat/}} \citep[$+46.53\pm0.12~{\rm{rad~m}}^{-2}$;][]{Manchester2005}.

\subsubsection{Pulsar candidates}
\cite{Lenc2017} also catalogue another pulsar (PSR~ B0740$-$28) in the region covered in this paper. With a measured FD of $+150.6\pm0.1$ rad~m$^{-2}$ and an integrated polarised flux density of 293~mJy~PSF$^{-1}$, it lies within the range considered in this work. However, this pulsar lies within the Galactic exclusion zone not covered by the GLEAM catalogue, so was not sampled as a candidate source. We will revisit the search for additional pulsars and candidates once the GLEAM Galactic catalogue becomes available.

Additionally, we searched our catalogue for further pulsar candidates, which would have to fulfil three criteria to be considered. Firstly, they must be compact at the resolution of both GLEAM and the NVSS. Secondly, they would have to exhibit a high polarisation fraction $(\gtrsim10\%)$. Thirdly, they would have to have steep radio spectra $(\alpha \lesssim -1)$. 

We found a single source that fulfilled these criteria: GLEAM~J134038$-$340234 (01-54 in Tables~\ref{tab:catalogue} and \ref{tab:class}). This source exhibits high polarisation fraction $m_{\rm{216~MHz}} = 23.3\pm7.1$ per cent. It also has a relatively large FD $-60.0\pm0.7$~rad~m$^{-2}$. We present a postage stamp image of this source, as well as the fitted spectral index, in Figure~\ref{fig:psrcand}. 

\begin{figure*}
\begin{center}
\includegraphics[width=0.7\textwidth]{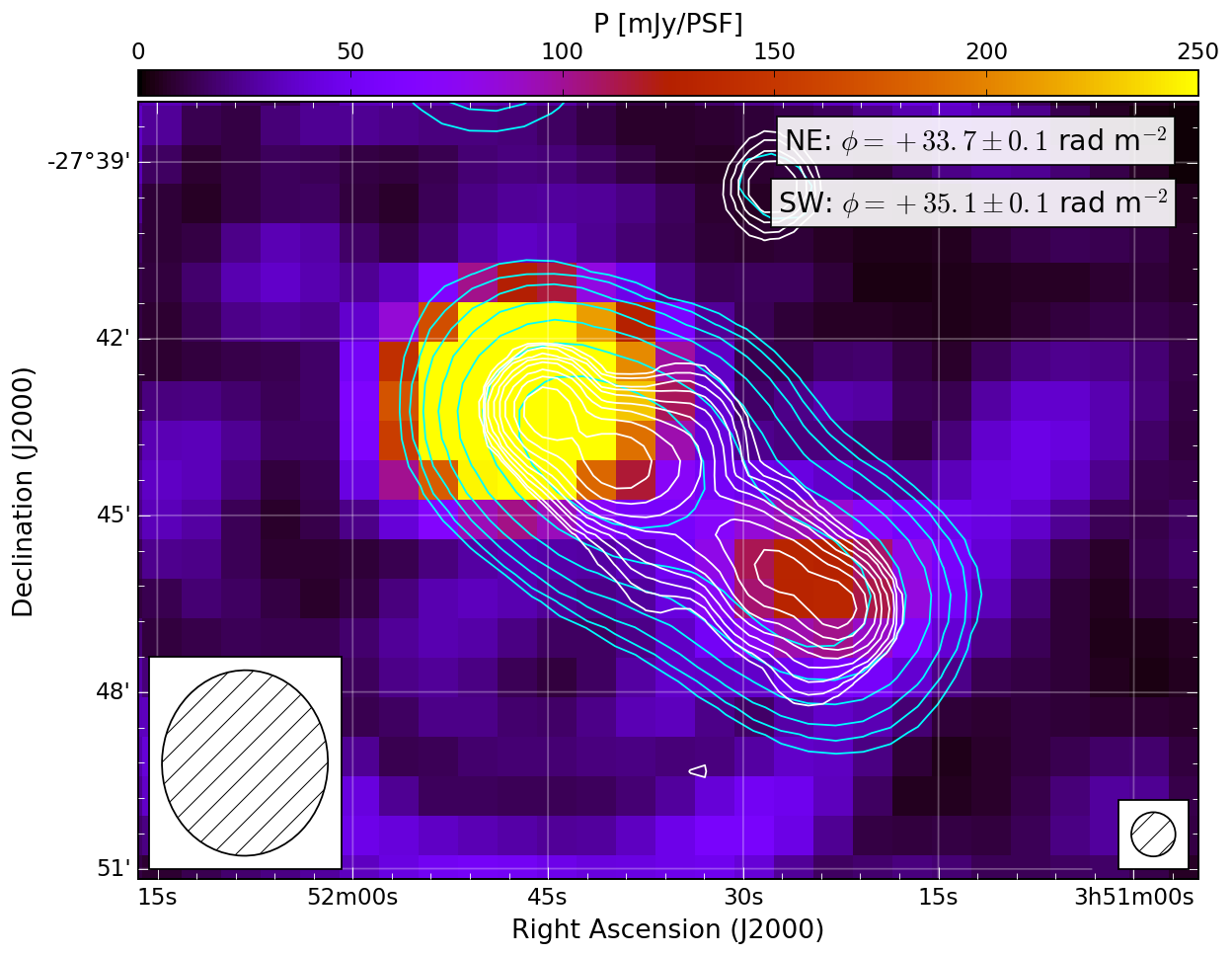} 
\caption{Linear polarisation image of PMN~J0351$-$2744, cut from a slice through our FD cube at $\phi=+34$ rad m$^{-2}$. Cyan contours are GLEAM continuum from the "white" mosaic (204--232~MHz) starting at $5\sigma_{\rm{local}}$ and scaling by a factor two. White contours are NVSS Stokes I, starting at 2.25~mJy PSF$^{-1}$ $(5\sigma)$ and scaling by a factor two. The hotspots are detected at slightly different FD, indicated in the inset. The relative PSF sizes of GLEAM and the NVSS are respectively indicated in the lower-left and lower-right corners.}
\label{fig:pmnj0351_postage}
\end{center}
\end{figure*}

Using the GLEAM flux density measurements in conjunction with those from the NVSS, Sydney University Molonglo Sky Survey \citep[SUMSS;][]{Mauch2003} and the Rescaled Subset of the first Alternative Data Release from the TIFR-GMRT Sky Survey \citep[TGSS-RSADR1;][]{HurleyWalker2017b,Intema2017} we find a best-fit power-law spectral index $\alpha=-1.2$. Whilst the relatively high local noise in GLEAM, combined with the faint flux density of this source, means that the GLEAM flux density measurements exhibit significant scatter, the observed trend is consistent with flux density measurements from the literature. This suggests this source as a pulsar candidate.

\subsection{Large-angular-scale radio sources}\label{src:lasrg}
Here we will discuss briefly the properties of the polarised emission from four LAS radio sources in our catalogue. For three of the LAS radio galaxies, there is good correspondence between the NVSS components and the GLEAM continuum morphology. As such, we re-derived the GLEAM 216~MHz flux densities with Aegean 2.0 \citep{Hancock2018} using the NVSS component positions as a prior. For ESO~422$-$G028, however, the correspondence is less clear, and we did not perform priorised fitting.

\subsubsection{PMN~J0351$-$2744}
The North-Eastern lobe of PMN~J0351$-$2744 (hereafter PMNJ0351) was the first polarised source detected by the MWA 32-tile prototype \citep{Bernardi2013}. Later, with the improved sensitivity and resolution afforded by the full 128-tile MWA, \cite{Lenc2017} detected polarised emission from both lobes. We present a postage stamp image of PMNJ0351 in Figure~\ref{fig:pmnj0351_postage}. We detect both lobes in polarisation, at slightly different FD, although both are consistent with previous MWA results.

PMNJ0351 is hosted by a Seyfert II galaxy at redshift $z=0.0656$ \citep{Mahony2011}. At this redshift, the angular diameter between hotspots (approx. 6.3~arcmin) corresponds to a physical separation of 0.49~Mpc.

\subsubsection{ESO~422$-$G028}
\begin{figure*}
\begin{center}
\includegraphics[width=0.85\textwidth]{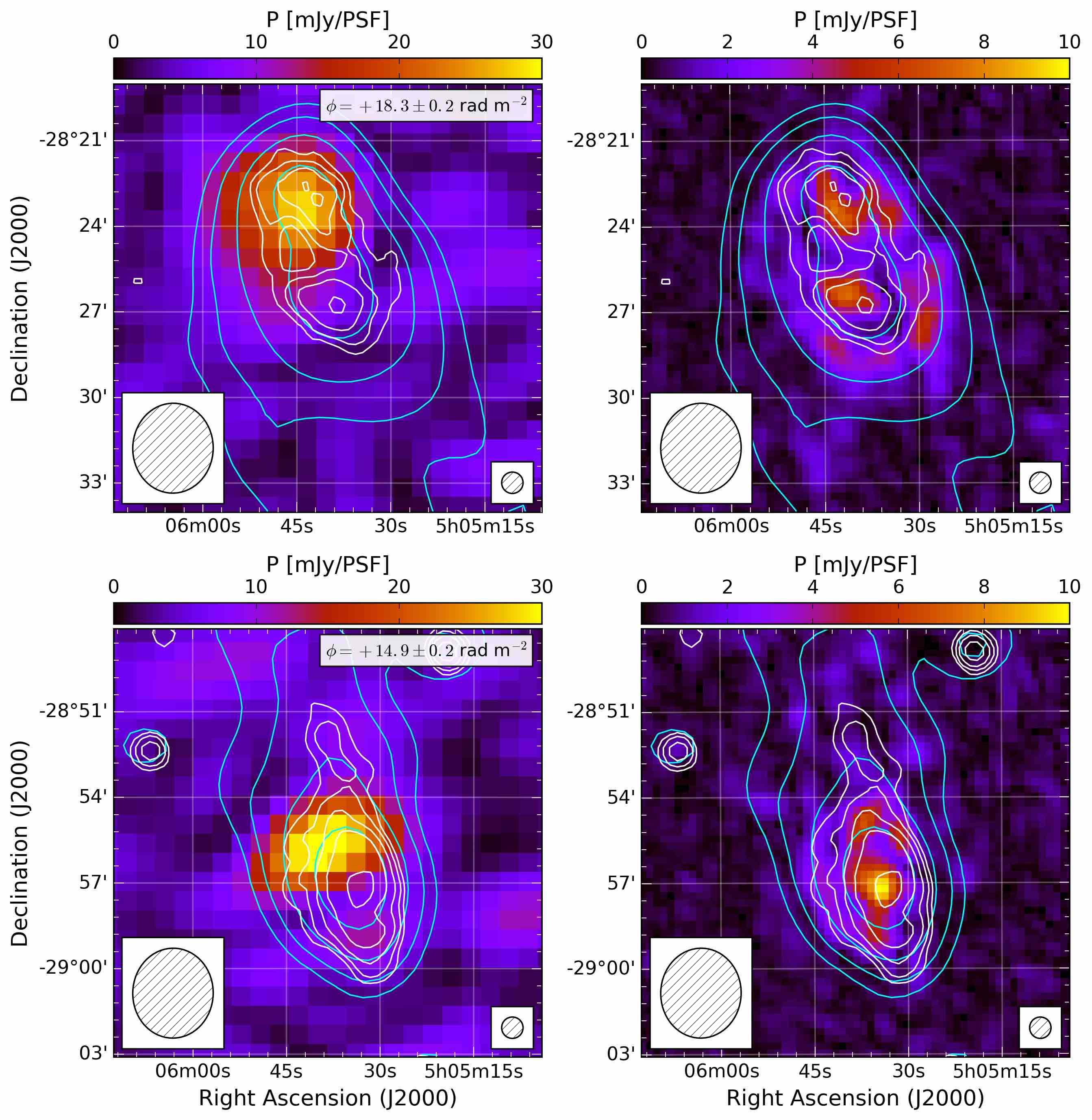}
\caption{Linear polarisation images of GLEAM~J050544$-$282236 (top) and GLEAM~J050535$-$285648 (bottom), the radio lobes of the giant radio galaxy ESO 422$-$G028. Left panels show the 216~MHz linearly-polarised intensity, sliced through our FD cube close to the fitted peak $(\phi=+18~{\rm{and}}~+15~{\rm{rad~m}}^{-2}$, respectively$)$. For each source, the fitted FD indicated in the inset. Right panels show 1.4~GHz linearly-polarised intensity from the NVSS. In both panels, cyan (white) contours are GLEAM (NVSS) Stokes I starting at $5\sigma$ and scaling by a factor two, as per Figure~\ref{fig:pmnj0351_postage}. Note that only the hotspot associated with the Southern radio lobe appears in the \protect\cite{Taylor2009} catalogue.}
\label{fig:gleamj0505_postage}
\end{center}
\end{figure*}

The highly extended radio lobes associated with this giant radio galaxy (GRG; also known in the literature as 0503$-$286) are hosted by the galaxy ESO~422$-$G028 (hereafter ESO~422) at redshift $z=0.0381$ \citep{Saripalli1986,Jamrozy2005}. At this redshift, the angular diameter between hotspots (approx. 34 arcmin) corresponds to a physical separation of 1.6~Mpc, given our cosmology. We present postage stamp images of the Northern and Southern lobes (respectively GLEAM~J050544$-$282236 and GLEAM~J050535$-$285648) in Figure~\ref{fig:gleamj0505_postage}. 

From Figure~\ref{fig:gleamj0505_postage}, we detect polarised emission from both lobes of this GRG. This emission i) appears at different FD and ii) occurs from starkly different regions. At 216~MHz, polarised emission in the North is associated with the hotspot, whereas in the South the polarised emission appears to be associated with a backflow region. 

In addition, the polarised emission at 1.4~GHz in the NVSS exhibits different morphology to that at 216~MHz. Whilst the Northern lobe is not catalogued by \cite{Taylor2009}, from Figure~\ref{fig:gleamj0505_postage} the 1.4~GHz emission to the North appears diffuse, with no obvious hotspot (in either total intensity or polarisation). This complex diffuse polarised structure is not detected by the MWA. A number of effects likely contribute to the observed morphology. Firstly, intrinsic turbulence in the lobe would disrupt the coherent magnetic field structure required for significant linear polarisation. Secondly, given our modest resolution at 216~MHz, the FD structure is likely tangled on scales smaller than our synthesised beam, leading to an observed depolarisation. Additionally, mixture between emitting and Faraday-rotating (relativistic and thermal) plasma would also lead to depolarisation \citep[see e.g.][for simulations of such structures in AGN lobes]{Hillel2016}.

Conversely, the Southern lobe exhibits an obvious hotspot, which is detected in polarisation by \cite{Condon1998}. As can be seen in Figure~\ref{fig:gleamj0505_postage}, the Southern lobe also hosts diffuse polarised emission at 1.4~GHz, with emission from the same region we detect at 216~MHz. However, no polarised emission is detected from the hotspot at 216~MHz.

The host, ESO~422 is a LINER galaxy \citep{Veron2010} with a large mass $(10^6 - 10^7~{\rm{M_{\odot}}})$ of warm $(35-50~{\rm{K}})$ dust \citep{Trifalenkov1994}, suggesting that the system is likely inclined with respect to the LOS. Given the asymmetry in the lobe morphology, we suggest that the Northern lobe is inclined toward the observer, with the Southern lobe pointing away. As such, it would almost certainly be subject to the Laing-Garrington effect \citep{Garrington1988,Laing1988}.

However, this would not explain the morphology of the polarised emission seen in the lower panel of Figure~\ref{fig:gleamj0505_postage}, as the source has a (projected) linear size of approx. 1.5~Mpc, and therefore extends well beyond the local medium of the host. Studying the physical cause of intrinsic depolarisation in radio galaxy lobes has become possible with the new generation of broad-band radio telescopes \citep[e.g.][]{OSullivan2018,Anderson2018}. In the case of the Southern hotspot of ESO422, there are two principal plausible explanations: hotspot complexes, or Kelvin-Helmholtz instabilities.

\paragraph{Hotspot complexes:}
It is now well-established that AGN hotspots often occur as `complexes' rather than a singular hotspot \citep[e.g.][]{Laing1982,Leahy1997,Carilli1999,Orienti2012}. These structures could arise as a result of episodic AGN activity, yielding vastly different polarimetric properties \citep[e.g.][]{Beuchert2018}. Such hotspot complexes may have superimposed Faraday structures (for example, these are visible in the Southern hotspot of PKS~J0636$-$2036 at 5~arcsec resolution; \citealt{OSullivan2018}) leading to the observed depolarisation at our moderate spatial resolution. Further high-resolution observations would be required to determine whether the structure of the Southern hotspot is more complex than it appears here.

\paragraph{Kelvin-Helmholtz instability:}
Kelvin-Helmholtz instabilities (KHI) arise as a result of velocity shear across the interface between radio lobes and a denser ambient medium. Given that the magnetic field is embedded in the plasma, the complex resulting magnetic field configurations \citep[e.g.][and references therein]{Ma2014} would lead to differential Faraday structure across the source \citep[e.g.][]{Bicknell1990,Sokoloff1998} induced by entrainment of thermal plasma from the ambient medium mixing with the relativistic plasma in the radio lobe.

Whilst modern MHD simulations do not possess the required numerical resolution to examine these effects on the surface of radio lobes, broadly speaking they suggest that the interface can generate KHI \citep[for example][]{HuarteEspinosa2011,Hardcastle2014}. However, this is also highly sensitive to the state of the AGN, the environment and the magnetic field \citep[see discussion by][and references therein]{Anderson2018}. Nevertheless, given the earlier evidence that we are viewing this system in projection, KHI entrainment of thermal plasma from the ambient medium remains a plausible explanation. Additional broad-band observations \citep[such as those presented by][]{OSullivan2018} would be required to investigate this further.

\subsubsection{PKS~J0636$-$2036}
\begin{figure*}
\begin{center}
\includegraphics[width=17.5cm]{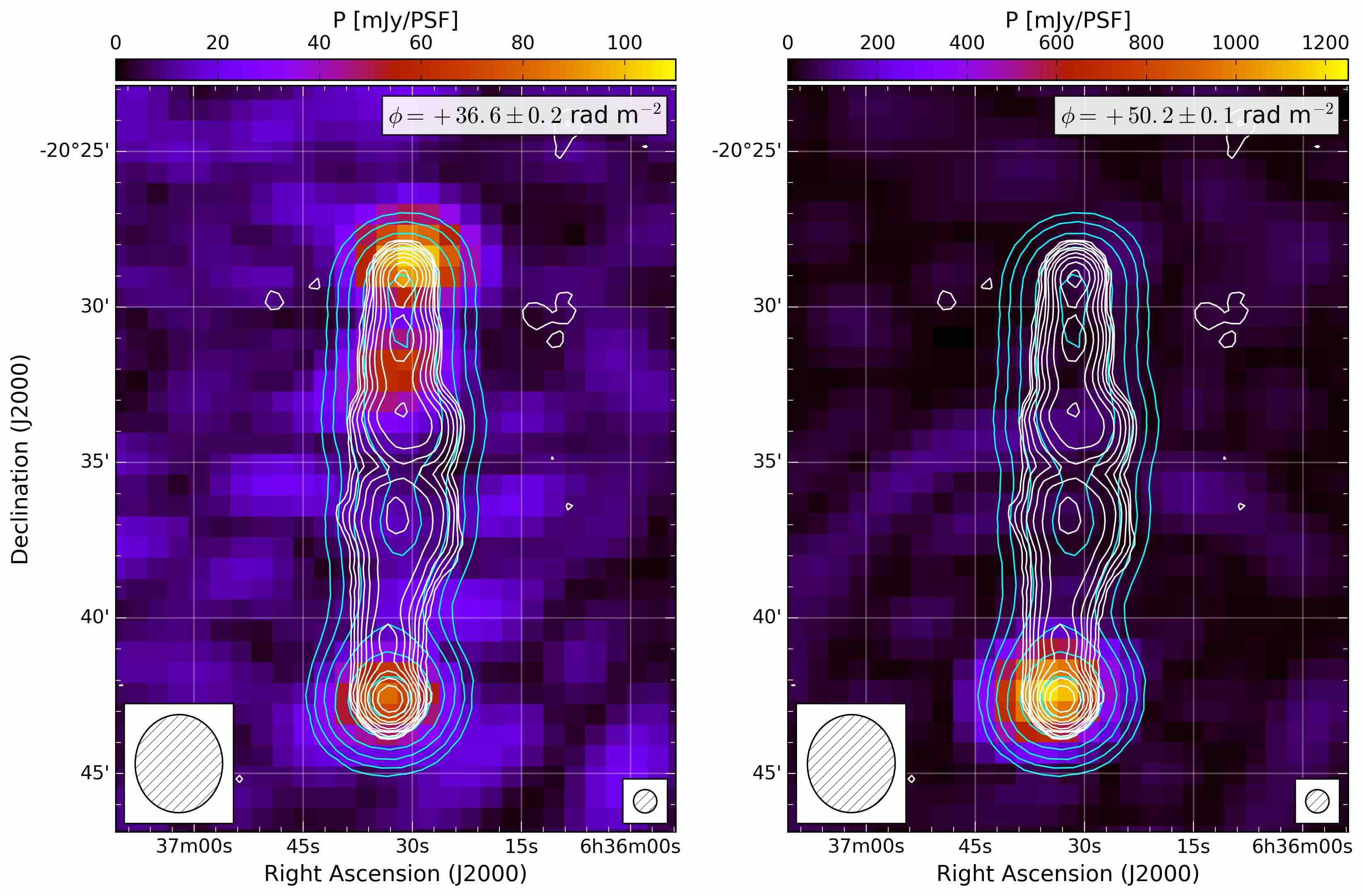} \\
\includegraphics[width=17.5cm]{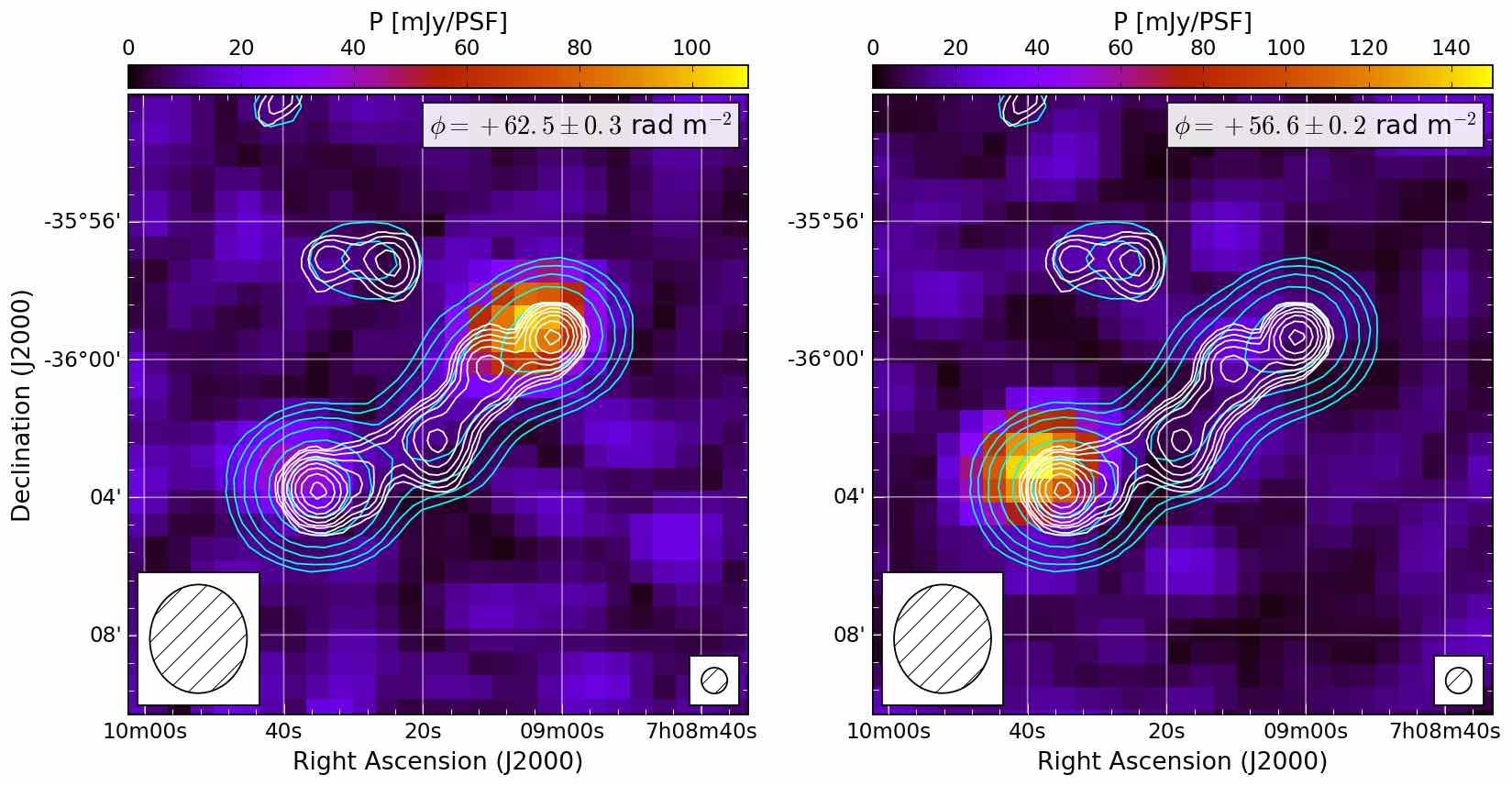} \\
\caption{Linear polarisation images of PKS~J0636$-$2036 (top) and PKS~0707$-$35 (bottom). Contours are as per Figure~\ref{fig:pmnj0351_postage}. Colourscales are cut from slices through our FD cubes close to the fitted peak. The fitted peak FD of each source is indicated in the inset. The relative PSF sizes of GLEAM and the NVSS are shown in the lower-left and lower-right corners, respectively.}\label{fig:doubles_postage}
\end{center}
\end{figure*}

We present postage stamp images of PKS~J0636$-$2036 (hereafter PKSJ0636) in the upper panels of Figure~\ref{fig:doubles_postage}. We detect two polarised sources associated with the Northern hotspot (NH) and Southern hotspot (SH) of this LAS radio galaxy. PKSJ0636 is hosted by an isolated elliptical galaxy at redshift $z=0.055$ \citep{McAdam1975}. At this redshift, the angular diameter between hotspots (approx. 14~arcmin) corresponds to a physical separation of 0.92~Mpc.

The SH is the brightest extragalactic polarised source yet detected at low frequencies, with a polarised flux density in our catalogue of $1.23\pm0.02~{\rm{Jy~PSF}}^{-1}$, and FD $\phi=+50.2\pm0.1$~rad~m$^{-2}$. We measure broadly consistent flux densities and FD to previous work with the MWA \citep[][]{Lenc2017,OSullivan2018}.

There is significant difference in the measured FD for the NH and SH: respectively $+36.6$ and $+50.2$~rad~m$^{-2}$. Whilst projection effects and/or asymmetries in the local environment may play some role in causing this difference, we consider it more likely that variations in Galactic FD on these scales (approx. 15~arcmin) are the dominant contribution to this difference. 

At present, this region of sky is too coarsely sampled to conclusively resolve this difference \citep[for example, the entire source lies within a single pixel of the Galactic FD map of][]{Oppermann2012,Oppermann2015}. However, many studies have probed the Galactic FD structure at significantly higher surface density across small regions of sky, either via statistical studies of many polarised sources \citep[e.g.][]{Haverkorn2004,Stil2011} or through FD variations in resolved polarised sources \citep[e.g.][]{Leahy1986,Simonetti1986,Minter1996}. These narrow-field studies have found significant variation in the Galactic FD across a wide range of angular scales, including the angular scale of PKSJ0636. Future highly-sensitive polarimetric surveys, such as the Polarisation Sky Survey of the Universe's Magnetism \citep[POSSUM;][]{Gaensler2010} with the Australian SKA Pathfinder \citep[ASKAP;][]{Johnston2007} will be crucial for improving our understanding of the Galactic magnetic field and FD.

The SH was catalogued by \cite{Taylor2009} with a similar RM ($+47.1\pm1.9$) to that measured by the MWA. However, we note that the NH was not catalogued by \cite{Taylor2009}; instead, they detect polarised emission associated with part of the Northern lobe. We also detect some emission from the same region (see the top-left panel of Figure~\ref{fig:doubles_postage}). However we could not conclusively exclude the possibility that it was associated with image-plane sidelobes from the SH, and thus we excluded this source from our catalogue. We refer the reader to \cite{OSullivan2018} for detailed modelling of the depolarisation mechanisms for the NH and SH of PKSJ0636.

\subsubsection{PKS~0707$-$35}\label{sec:pks0707}
PKS~0707$-35$ (hereafter PKS0707) is hosted by a Seyfert II galaxy at redshift $z=0.11$ \citep{Burgess2006}. At this redshift, the angular diameter between hotspots (approx. 8.5 arcmin) corresponds to a physical separation of 1.1~Mpc. We present postage stamp images of PKS0707 in the lower panels of Figure~\ref{fig:doubles_postage}. 

We detect two polarised sources associated with the hotspots of PKS0707. For the NW (SE) lobe, our measured FD are $\phi=+62.5\pm0.3$ $(\phi=+56.6\pm0.2)$ rad~m$^{-2}$. The NW hotspot is consistent with the NVSS RM: $\phi_{\rm{NVSS}} = +56.5\pm12.6$, although the SE hotspot exhibits a discrepancy $(\phi_{\rm{NVSS}} = +40.3\pm9.1)$ rad~m$^{-2}$. 

Additionally, both hotspots exhibit slight repolarisation, with a depolarisation ratio $m_{\rm{216MHz}}/m_{\rm{1.4GHz}} = 1.34$ for the SE hotspot and $m_{\rm{216MHz}}/m_{\rm{1.4GHz}} = 1.14$ for the NW hotspot. This suggests that these hotspots are both `Faraday-complex'. Further broad-band polarimetric study is required to further understand the Faraday structure of this GRG. Additional observations at higher resolution are also required to investigate whether this repolarisation is caused by multiple unresolved components within our synthesised beam \citep[e.g.][]{Farnes2014b}.

\subsubsection{Discussion}
There are no CD sources where both hotspots exhibit polarised emission. One explanation for this would be the Laing-Garrington effect \citep{Garrington1988,Laing1988}; whereby the more distant lobe is depolarised by the medium local to the host galaxy. However, we lack sufficient information on the source orientation with respect to the plane of the sky to test this hypothesis.

Conversely, the majority of ED sources in our catalogue have two polarised detections, typically corresponding to AGN hotspots. We know that these ED sources exist on scales in excess of several hundred kpc, far beyond the local environment of the host galaxy, and as such should not be affected by internal Faraday depolarisation. However, as we have discussed, the Southern hotspot of ESO422 is depolarised, likely due to complex Faraday structure in the lobe region. 

In the unified AGN scheme, it should be possible to use these ED sources as analogues to CD sources at higher redshift. However, a number of cases this is likely not possible. Consider the CD source GLEAM~J215657$-$181343\footnote{01-75 in Figure~\ref{fig:multi_postage} and Tables~\ref{tab:catalogue} and \ref{tab:class}.} (which has a single polarised component) hosted by a QSO at redshift $z=0.668$ \citep{Flesch2017}. At this redshift, the angular diameter between hotspots (approx. 1.23~arcmin) corresponds to a physical separation of 0.53~Mpc. While no information is available on the host viewing angle, this is in excess of the projected distance between the hotspots of PMNJ0351 (0.49~Mpc) which has two polarised components. 

Furthermore, there are three LAS radio galaxies in our catalogue (01-03, 01-04 and 01-79) which only have a single polarised component. All have identified host galaxies with measured spectroscopic redshifts. For all three, the projected physical separation between hotspots (respectively 0.73~Mpc, 0.51~Mpc and 0.56~Mpc) are also greater than that of PMNJ0351.

\section{CONCLUSIONS AND FUTURE PROSPECTS}
In this paper, we have presented the first results from the POGS project. We have re-processed a sample of GLEAM drift-scan observations centred on Declination $-27\degree$ (zenith from the MWA site) covering 24 hours in Right Ascension, with the aim of characterising the low-frequency linearly-polarised source population. We have exploited recent advances in polarimetric data processing as well as source-finding and characterisation in polarimetric surveys. We have catalogued 81 sources (including a known pulsar and new pulsar candidate) above a threshold of 18~mJy PSF$^{-1}$ across a region of 6400 square degrees. This implies a sky surface density of 0.013~deg$^{-2}$, or one source per 79 square degrees. Assuming this is representative of the entire sky, we predict we should detect in excess of 300 sources in the full POGS catalogue.

Our catalogue contains 71 sources in common with the 1.4~GHz NVSS rotation measure catalogue \citep{Taylor2009}. The Faraday depths recovered for our POGS catalogue show general correspondence with their counterparts at 1.4~GHz, although some sources show a difference that is too large to be explained solely by measurement uncertainties. This suggests that a sub-sample of sources may either have some Faraday depth structure too broad to be probed by the MWA and/or comprised multiple components with differing polarisation and spectral properties. 

Additionally, we have performed an initial analysis of depolarisation between 216~MHz and 1.4~GHz. The majority of sources exhibit a reduced polarisation fraction at 216~MHz. However, a sub-sample appear to repolarise, perhaps suggesting multiple emission regions (with different polarisation properties and spectral indices) internal to these sources. 

We have cross-referenced our catalogue with the literature in order to investigate the nature of these polarised sources. We were able to associate approximately two thirds of the sources in our catalogue with a host object; with one exception (the pulsar PSR~B0628$-$28) all are associated with AGN-type objects. This suggests that the majority of polarised emission seen by the MWA originates from highly-ordered magnetic fields in the termination hotspots of radio jets. Based on spectral, morphological and polarisation properties, we also identify a new pulsar candidate among those sources without identified host sources.

A handful of sources in our catalogue are associated with nearby, LAS radio sources, which we have discussed in detail. In particular, we have focussed on the GRG ESO~422$-$G028, attempting to explain the striking difference in the morphology of the polarised radio emission detected at 216~MHz and 1.4~GHz. Given what is known about the host galaxy, we suggest that the complex Faraday structure implied by the depolarisation seen in the Southern radio lobe is the result of either multiple hotspots or Kelvin-Helmholtz instabilities. Wide-band followup at higher resolution would be required to discriminate between these scenarios.

\subsection{Future Work}\label{sec:future}
\subsubsection{Broad-band polarimetry}
In this paper, we considered a single GLEAM band covering $204-232$~MHz, covering a single strip $20\degree$ wide, centred on Declination~$-27\degree$. Paper II in this series (Riseley et al., in prep.) will expand on both the sky coverage as well as the frequency range explored.

Further ahead, by taking advantage of more of the MWA bandwidth, in conjunction with upcoming ASKAP \citep{Johnston2007} surveys, such as POSSUM \citep[which will cover the frequency range $1130-1430$~MHz;][]{Gaensler2010}, we will be able to construct a unique all-sky sample of polarised radio sources. POGS and POSSUM will be highly complementary, yielding a product with both short $\lambda^2$ spacings and high Faraday-space resolution. The extreme fractional bandwidth provided by such a sample will enable us to model the depolarisation behaviour over a broad $\lambda^2$ range to provide strong constraints on the magnetoionic structure of AGN lobes \citep[e.g.][]{OSullivan2018}.

\subsubsection{Extended MWA configuration}
Additionally, the MWA has recently completed its Phase II expansion \citep{Wayth2018}. This has approximately doubled the maximum baseline to 6~km; at the frequencies considered in this work, it should be possible to achieve resolution of the order of an arcminute. A number of wide and deep surveys are planned with the extended MWA; polarisation processing of these data should yield an expanded catalogue due to improvements in both sensitivity and resolution, the latter of which will help mitigate beam depolarisation.

\begin{acknowledgements}
This work makes use of the Murchison Radioastronomy Observatory, operated by CSIRO. We acknowledge the Wajarri Yamatji people as the traditional owners of the Observatory site. Support for the operation of the MWA is provided by the Australian Government (NCRIS) under a contract to Curtin University, administered by Astronomy Australia Limited. This work was supported by resources provided by the Pawsey Supercomputing Centre, with funding from the Australian Government and the Government of Western Australia.  

We thank our anonymous referee for their thorough review of our paper, and their comments which have contributed to the improvement of this manuscript. The Dunlap Institute is funded through an endowment established by the David Dunlap family and the University of Toronto. B.M.G. acknowledges the support of the Natural Sciences and Engineering Research Council of Canada (NSERC) through grant RGPIN-2015-05948, and of the Canada Research Chairs program. We acknowledge the International Centre for Radio Astronomy Research (ICRAR), which is a joint venture between Curtin University and The University of Western Australia, funded by the Western Australian State government.

This research has made use of NASA's Astrophysics Data System (ADS) and the NASA/IPAC Extragalactic Database (NED) as well as the VizieR catalogue access tool, CDS, Strasbourg, France. This research used ionospheric TEC maps produced by the Centre for Orbital Determination
in Europe (CODE, \url{http://aiuws.unibe.ch/ionosphere/}). We also wish to acknowledge the developers of the following python packages, which were used during this work: \texttt{aplpy} \citep{Robitaille2012}, \texttt{astropy} \citep{Astropy2013}, \texttt{lmfit} \citep{Newville2017}, \texttt{matplotlib} \citep{Hunter2007}, \texttt{numpy} \citep{Numpy2011} and \texttt{scipy} \citep{Jones2001}.

\end{acknowledgements}

\bibliographystyle{pasa-mnras}
\bibliography{POGS-I}

\appendix
\section{Catalogue of linearly-polarised POGS sources}

\onecolumn
\begin{landscape}
{\footnotesize
\begin{longtable}{p{1cm}lp{1.5cm}p{1.5cm}|rrrr|rr|l}
\caption{Catalogue of linearly-polarised point sources detected in the Dec$-27\degree$ strip. Note that we cannot extract a specific ionospheric FD for each source in our catalogue, as the ionospheric correction was applied on a per-snapshot basis, whilst our source-finding was performed on our mosaicked FD cubes. \label{tab:catalogue}}\\
\hline
& & \multicolumn{2}{c}{Polarized peak coordinates} & \multicolumn{4}{|c|}{POGS} & \multicolumn{2}{|c|}{NVSS$^{\ddag}$} & \\
POGS ID & GLEAM ID   & Right Ascension   & Declination     &  $S_{\rm{I}}^{\ast}$  &  $P$ 
         & $m^{\dag}$ & RM   
         & $m$ & RM  & Notes$^{\ast}$ \\
&          & (J2000)        & (J2000)          & [Jy]                  & [mJy/PSF] 
         & [$\%$]         &               [rad m$^{-2}$]
         & [$\%$]           & [rad m$^{-2}$]  & \\
\hline
\endfirsthead
\caption{(continued).}\\
\hline
& & \multicolumn{2}{c}{Polarized peak coordinates} & \multicolumn{4}{|c|}{POGS} & \multicolumn{2}{|c|}{NVSS$^{\ddag}$} & \\
POGS ID & GLEAM ID   & Right Ascension   & Declination     &  $S_{\rm{I}}^{\ast}$  &  $P$ 
         & $m^{\dag}$ & RM   
         & $m$ & RM  & Notes$^{\ast}$ \\
&          & (J2000)        & (J2000)          & [Jy]                  & [mJy/PSF] 
         & [$\%$]         &               [rad m$^{-2}$]
         & [$\%$]           & [rad m$^{-2}$]  & \\
\hline
\endhead
\hline
\multicolumn{11}{l}{$^\dag$: Fractional polarisation ($P$/$I$). $\ddag$: From the \protect\citet{Taylor2009} catalogue; a value of "$-$" denotes no counterpart. $^\ast$: Stokes I flux density measurement: `G' indicates that $S_{\rm{I}}$}\\ 
\multicolumn{11}{l}{is the 212~MHz measurement from the GLEAM catalogue \citep{HurleyWalker2017}, `P' indicates a measurement derived by priorised fitting with Aegean \citep{Hancock2018}, }\\
\multicolumn{11}{l}{where the polarised emission is clearly associated with a single NVSS component, and `G*' indicates a source where the polarised emission is not clearly associated with a single}\\
\multicolumn{11}{l}{NVSS component. In these situations, we used the 212~MHz GLEAM flux density measurement.}
\endfoot
\hline
\multicolumn{11}{l}{$^\dag$: Fractional polarisation ($P$/$I$). $\ddag$: From the \protect\citet{Taylor2009} catalogue; a value of "$-$" denotes no counterpart. $^\ast$: Stokes I flux density measurement: `G' indicates that $S_{\rm{I}}$}\\ 
\multicolumn{11}{l}{is the 212~MHz measurement from the GLEAM catalogue \citep{HurleyWalker2017}, `P' indicates a measurement derived by priorised fitting with Aegean \citep{Hancock2018}, }\\
\multicolumn{11}{l}{where the polarised emission is clearly associated with a single NVSS component, and `G*' indicates a source where the polarised emission is not clearly associated with a single}\\
\multicolumn{11}{l}{NVSS component. In these situations, we used the 212~MHz GLEAM flux density measurement.}
\endlastfoot
\hline
01-01 & GLEAM J001341$-$300925 & 003.42234 & $ -30.15570 $ & $ 0.81 \pm 0.02 $ & $ 21.8 \pm 8.1 $ & $ 2.7 \pm 1.0 $ & $ -9.2 \pm 1.3 $ & $ - $ & $ - $ & G  \\
01-02 & GLEAM J002021$-$202846 & 005.09602 & $ -20.47940 $ & $ 1.13 \pm 0.01 $ & $ 58.2 \pm 3.9 $ & $ 5.2 \pm 0.3 $ & $ -7.3 \pm 0.3 $ & $ 8.57 \pm 0.15 $ & $ -11.5 \pm 4.8 $ & P  \\
01-03 & GLEAM J002026$-$201427 & 005.11111 & $ -20.23090 $ & $ 2.19 \pm 0.01 $ & $ 39.5 \pm 6.6 $ & $ 1.8 \pm 0.3 $ & $ -20.8 \pm 0.6 $ & $ 13.74 \pm 0.15 $ & $ 1.5 \pm 2.7 $ & P  \\
01-04 & GLEAM J002112$-$191041 & 005.31153 & $ -19.17720 $ & $ 3.90 \pm 0.01 $ & $ 150.3 \pm 7.8 $ & $ 3.9 \pm 0.2 $ & $ 9.0 \pm 0.2 $ & $ 3.67 \pm 0.07 $ & $ 3.6 \pm 5.2 $ & P  \\
01-05 & GLEAM J003906$-$242506 & 009.78630 & $ -24.41370 $ & $ 0.99 \pm 0.01 $ & $ 29.5 \pm 3.8 $ & $ 3.0 \pm 0.4 $ & $ 15.8 \pm 0.4 $ & $ 5.14 \pm 0.25 $ & $ 6.1 \pm 9.8 $ & G  \\
01-06 & GLEAM J014709$-$223234 & 026.79727 & $ -22.54620 $ & $ 1.58 \pm 0.01 $ & $ 33.9 \pm 6.5 $ & $ 2.1 \pm 0.4 $ & $ 13.1 \pm 0.7 $ & $ - $ & $ - $ & G  \\
01-07 & GLEAM J021726$-$224207 & 034.35776 & $ -22.70040 $ & $ 0.41 \pm 0.01 $ & $ 26.1 \pm 4.6 $ & $ 6.4 \pm 1.1 $ & $ 13.1 \pm 0.6 $ & $ 8.77 \pm 0.43 $ & $ 32.5 \pm 10.4 $ & G  \\
01-08 & GLEAM J023512$-$293622 & 038.79565 & $ -29.60500 $ & $ 0.67 \pm 0.01 $ & $ 45.8 \pm 2.2 $ & $ 6.8 \pm 0.3 $ & $ 7.8 \pm 0.2 $ & $ 5.95 \pm 0.21 $ & $ 9.9 \pm 7.5 $ & G*  \\
01-09 & GLEAM J023700$-$283448 & 039.24998 & $ -28.57160 $ & $ 0.69 \pm 0.01 $ & $ 35.4 \pm 2.5 $ & $ 5.1 \pm 0.4 $ & $ 9.0 \pm 0.3 $ & $ 19.73 \pm 0.33 $ & $ 5.3 \pm 2.4 $ & P  \\
01-10 & GLEAM J023845$-$223321 & 039.70466 & $ -22.55240 $ & $ 0.90 \pm 0.01 $ & $ 22.3 \pm 4.0 $ & $ 2.5 \pm 0.5 $ & $ 13.4 \pm 0.5 $ & $ 4.22 \pm 0.26 $ & $ 7.1 \pm 13.1 $ & G*  \\
01-11 & GLEAM J031152$-$312959 & 047.97066 & $ -31.49710 $ & $ 2.68 \pm 0.01 $ & $ 141.4 \pm 4.8 $ & $ 5.3 \pm 0.2 $ & $ 14.4 \pm 0.1 $ & $ 15.11 \pm 0.36 $ & $ 24.7 \pm 5.9 $ & P  \\
01-12 & GLEAM J032022$-$312458 & 050.09630 & $ -31.41260 $ & $ 2.49 \pm 0.01 $ & $ 43.3 \pm 9.4 $ & $ 1.7 \pm 0.4 $ & $ 22.9 \pm 0.7 $ & $ 4.94 \pm 0.06 $ & $ 20.6 \pm 2.7 $ & G  \\
01-13 & GLEAM J032800$-$220205 & 052.00807 & $ -22.03530 $ & $ 0.73 \pm 0.01 $ & $ 24.0 \pm 2.7 $ & $ 3.3 \pm 0.4 $ & $ 23.5 \pm 0.4 $ & $ 4.83 \pm 0.07 $ & $ 28.1 \pm 2.9 $ & G  \\
01-14 & GLEAM J033810$-$250213 & 054.55624 & $ -25.03750 $ & $ 0.38 \pm 0.01 $ & $ 18.7 \pm 2.0 $ & $ 4.9 \pm 0.5 $ & $ 29.8 \pm 0.4 $ & $ 7.78 \pm 0.38 $ & $ 42.1 \pm 9.1 $ & G  \\
01-15 & GLEAM J034850$-$294359 & 057.21690 & $ -29.73370 $ & $ 1.32 \pm 0.01 $ & $ 27.1 \pm 3.3 $ & $ 2.0 \pm 0.3 $ & $ 14.0 \pm 0.1 $ & $ 4.51 \pm 0.16 $ & $ 44.8 \pm 7.4 $ & P  \\
01-16 & GLEAM J035125$-$274610 & 057.86545 & $ -27.76010 $ & $ 1.27 \pm 0.02 $ & $ 132.1 \pm 5.8 $ & $ 10.4 \pm 0.5 $ & $ 35.1 \pm 0.1 $ & $ - $ & $ - $ & P  \\
01-17 & GLEAM J035140$-$274354 & 057.94203 & $ -27.71580 $ & $ 4.18 \pm 0.02 $ & $ 591.3 \pm 15.6 $ & $ 14.2 \pm 0.4 $ & $ 33.7 \pm 0.1 $ & $ 20.06 \pm 0.52 $ & $ 34.7 \pm 5.5 $ & P  \\
01-18 & GLEAM J035537$-$203324 & 058.91729 & $ -20.55160 $ & $ 1.11 \pm 0.01 $ & $ 39.1 \pm 3.6 $ & $ 3.5 \pm 0.3 $ & $ 24.9 \pm 0.3 $ & $ 11.58 \pm 0.08 $ & $ 27.6 \pm 1.4 $ & G  \\
01-19 & GLEAM J044119$-$295231 & 070.32687 & $ -29.87060 $ & $ 1.46 \pm 0.01 $ & $ 18.7 \pm 4.0 $ & $ 1.3 \pm 0.3 $ & $ 29.6 \pm 1.0 $ & $ 1.06 \pm 0.06 $ & $ 58.3 \pm 11.7 $ & G  \\
01-20 & GLEAM J045714$-$270436 & 074.31943 & $ -27.07330 $ & $ 0.35 \pm 0.01 $ & $ 24.1 \pm 2.6 $ & $ 6.9 \pm 0.8 $ & $ 26.1 \pm 0.3 $ & $ 3.02 \pm 0.19 $ & $ 23.7 \pm 12.9 $ & G  \\
01-21 & GLEAM J050535$-$285648 & 076.40995 & $ -28.92670 $ & $ 2.48 \pm 0.03 $ & $ 34.8 \pm 2.1 $ & $ 1.4 \pm 0.1 $ & $ 14.9 \pm 0.2 $ & $ 19.52 \pm 1.13 $ & $ 19.9 \pm 14.7 $ & G  \\
01-22 & GLEAM J050544$-$282236 & 076.43958 & $ -28.39230 $ & $ 1.16 \pm 0.02 $ & $ 29.4 \pm 2.5 $ & $ 2.5 \pm 0.2 $ & $ 18.3 \pm 0.2 $ & $ - $ & $ - $ & G  \\
01-23 & GLEAM J050922$-$282427 & 077.34909 & $ -28.41000 $ & $ 0.83 \pm 0.01 $ & $ 22.5 \pm 2.0 $ & $ 2.7 \pm 0.2 $ & $ 15.9 \pm 0.3 $ & $ 17.36 \pm 0.34 $ & $ 13.1 \pm 3.3 $ & G  \\
01-24 & GLEAM J051024$-$195950 & 077.61214 & $ -19.99460 $ & $ 1.29 \pm 0.01 $ & $ 44.4 \pm 3.5 $ & $ 3.4 \pm 0.3 $ & $ 11.8 \pm 0.4 $ & $ 5.94 \pm 0.09 $ & $ 22.6 \pm 3.3 $ & G  \\
01-25 & GLEAM J051836$-$240305 & 079.67405 & $ -24.05500 $ & $ 0.37 \pm 0.01 $ & $ 43.8 \pm 2.7 $ & $ 11.8 \pm 0.8 $ & $ 29.4 \pm 0.2 $ & $ 8.21 \pm 0.33 $ & $ 43.7 \pm 8.7 $ & P  \\
01-26 & GLEAM J052618$-$260409 & 081.58587 & $ -26.06820 $ & $ 0.31 \pm 0.01 $ & $ 20.9 \pm 2.9 $ & $ 6.7 \pm 0.9 $ & $ 17.5 \pm 0.4 $ & $ 9.41 \pm 0.65 $ & $ 13.1 \pm 11.8 $ & G  \\
01-27 & GLEAM J055222$-$261015 & 088.09782 & $ -26.16310 $ & $ 1.04 \pm 0.01 $ & $ 18.0 \pm 3.4 $ & $ 1.7 \pm 0.3 $ & $ 13.3 \pm 0.9 $ & $ 7.43 \pm 0.14 $ & $ 38.7 \pm 3.9 $ & G  \\
01-28 & GLEAM J060753$-$263806 & 091.98111 & $ -26.63210 $ & $ 0.26 \pm 0.01 $ & $ 22.5 \pm 3.2 $ & $ 8.7 \pm 1.3 $ & $ 27.5 \pm 0.4 $ & $ 12.92 \pm 0.63 $ & $ 29.9 \pm 10.5 $ & P  \\
01-29 & GLEAM J060845$-$271708 & 092.19164 & $ -27.27440 $ & $ 0.45 \pm 0.01 $ & $ 22.3 \pm 2.7 $ & $ 5.0 \pm 0.6 $ & $ 23.2 \pm 0.3 $ & $ 5.19 \pm 0.12 $ & $ 25.8 \pm 4.8 $ & G  \\
01-30 & GLEAM J062126$-$210059 & 095.36862 & $ -21.01870 $ & $ 0.59 \pm 0.01 $ & $ 42.4 \pm 4.3 $ & $ 7.2 \pm 0.7 $ & $ 46.7 \pm 0.3 $ & $ 8.85 \pm 0.11 $ & $ 14.0 \pm 2.9 $ & G  \\
01-31 & GLEAM J062702$-$241925 & 096.75862 & $ -24.32350 $ & $ 0.31 \pm 0.01 $ & $ 38.7 \pm 3.4 $ & $ 12.5 \pm 1.2 $ & $ 75.9 \pm 0.3 $ & $ 11.22 \pm 0.58 $ & $ 65.5 \pm 12.4 $ & G  \\
01-32 & GLEAM J063049$-$283438 & 097.70592 & $ -28.57450 $ & $ 0.39 \pm 0.01 $ & $ 182.9 \pm 4.2 $ & $ 46.9 \pm 1.6 $ & $ 46.6 \pm 0.1 $ & $ 31.72 \pm 2.19 $ & $ 15.8 \pm 16.1 $ & G  \\
01-33 & GLEAM J063228$-$272109 & 098.11735 & $ -27.35030 $ & $ 2.20 \pm 0.01 $ & $ 53.2 \pm 4.3 $ & $ 2.4 \pm 0.2 $ & $ 111.1 \pm 0.3 $ & $ 13.65 \pm 0.33 $ & $ 108.0 \pm 3.5 $ & P  \\
01-34 & GLEAM J063631$-$202924 & 099.13146 & $ -20.48250 $ & $ 3.85 \pm 0.01 $ & $ 111.8 \pm 7.0 $ & $ 2.9 \pm 0.2 $ & $ 36.6 \pm 0.2 $ & $ 14.67 \pm 0.59 $ & $ 34.8 \pm 7.2 $ & P  \\
01-35 & GLEAM J063633$-$204225 & 099.13957 & $ -20.70710 $ & $ 11.02 \pm 0.01 $ & $ 1230.2 \pm 24.6 $ & $ 11.2 \pm 0.2 $ & $ 50.2 \pm 0.1 $ & $ 15.48 \pm 0.17 $ & $ 47.1 \pm 1.9 $ & P  \\
01-36 & GLEAM J065716$-$320328 & 104.32034 & $ -32.05650 $ & $ 3.35 \pm 0.01 $ & $ 45.5 \pm 7.8 $ & $ 1.4 \pm 0.2 $ & $ 54.2 \pm 0.4 $ & $ 5.32 \pm 0.02 $ & $ 57.5 \pm 1.0 $ & G  \\
01-37 & GLEAM J070901$-$355921 & 107.26303 & $ -35.98380 $ & $ 2.13 \pm 0.01 $ & $ 107.3 \pm 8.0 $ & $ 5.0 \pm 0.4 $ & $ 62.5 \pm 0.3 $ & $ 4.41 \pm 0.28 $ & $ 56.5 \pm 12.6 $ & P  \\
01-38 & GLEAM J070934$-$360341 & 107.40675 & $ -36.05480 $ & $ 4.05 \pm 0.01 $ & $ 171.7 \pm 8.2 $ & $ 4.2 \pm 0.2 $ & $ 56.6 \pm 0.2 $ & $ 3.15 \pm 0.12 $ & $ 40.3 \pm 9.1 $ & P  \\
01-39 & GLEAM J091245$-$251254 & 138.19762 & $ -25.21320 $ & $ 1.84 \pm 0.01 $ & $ 28.1 \pm 4.7 $ & $ 1.5 \pm 0.3 $ & $ -92.5 \pm 0.4 $ & $ 16.24 \pm 0.25 $ & $ -91.5 \pm 3.0 $ & G*  \\
01-40 & GLEAM J092317$-$213744 & 140.82271 & $ -21.62390 $ & $ 1.48 \pm 0.01 $ & $ 49.4 \pm 4.1 $ & $ 3.3 \pm 0.3 $ & $ -141.3 \pm 0.3 $ & $ 1.56 \pm 0.10 $ & $ -85.5 \pm 13.8 $ & P  \\
01-41 & GLEAM J095750$-$283808 & 149.44958 & $ -28.63490 $ & $ 1.44 \pm 0.01 $ & $ 30.6 \pm 5.8 $ & $ 2.1 \pm 0.4 $ & $ 138.3 \pm 0.6 $ & $ 6.71 \pm 0.20 $ & $ 155.8 \pm 6.7 $ & P  \\
01-42 & GLEAM J100055$-$193108 & 150.24382 & $ -19.51670 $ & $ 1.16 \pm 0.02 $ & $ 62.0 \pm 4.7 $ & $ 5.3 \pm 0.4 $ & $ 16.4 \pm 0.3 $ & $ 6.08 \pm 0.17 $ & $ 11.7 \pm 6.2 $ & G  \\
01-43 & GLEAM J100206$-$265606 & 150.52546 & $ -26.92660 $ & $ 0.59 \pm 0.01 $ & $ 26.5 \pm 4.5 $ & $ 4.5 \pm 0.8 $ & $ 152.1 \pm 0.4 $ & $ 8.37 \pm 0.16 $ & $ 138.7 \pm 4.2 $ & G  \\
01-44 & GLEAM J102056$-$321100 & 155.24853 & $ -32.17510 $ & $ 3.10 \pm 0.01 $ & $ 47.5 \pm 6.8 $ & $ 1.5 \pm 0.2 $ & $ -75.3 \pm 0.4 $ & $ 4.96 \pm 0.08 $ & $ -81.5 \pm 3.9 $ & G  \\
01-45 & GLEAM J110331$-$325116 & 165.88266 & $ -32.85710 $ & $ 3.18 \pm 0.01 $ & $ 52.5 \pm 6.2 $ & $ 1.6 \pm 0.2 $ & $ 37.5 \pm 0.4 $ & $ 2.68 \pm 0.08 $ & $ 16.2 \pm 7.3 $ & G  \\
01-46 & GLEAM J111007$-$272960 & 167.52723 & $ -27.49540 $ & $ 1.45 \pm 0.01 $ & $ 25.6 \pm 4.5 $ & $ 1.8 \pm 0.3 $ & $ -12.5 \pm 0.5 $ & $ 1.33 \pm 0.09 $ & $ 43.2 \pm 14.4 $ & G  \\
01-47 & GLEAM J120533$-$263407 & 181.38474 & $ -26.56650 $ & $ 4.73 \pm 0.01 $ & $ 44.3 \pm 8.1 $ & $ 0.9 \pm 0.2 $ & $ -35.1 \pm 0.5 $ & $ 1.11 \pm 0.04 $ & $ -149.5 \pm 12.8 $ & G  \\
01-48 & GLEAM J122646$-$324711 & 186.68969 & $ -32.78070 $ & $ 1.56 \pm 0.01 $ & $ 44.1 \pm 8.8 $ & $ 2.8 \pm 0.6 $ & $ -55.6 \pm 0.6 $ & $ 2.68 \pm 0.10 $ & $ -52.6 \pm 6.6 $ & G  \\
01-49 & GLEAM J130025$-$231806 & 195.10481 & $ -23.30730 $ & $ 4.82 \pm 0.02 $ & $ 116.0 \pm 4.3 $ & $ 2.4 \pm 0.1 $ & $ -25.0 \pm 0.1 $ & $ 2.66 \pm 0.05 $ & $ -30.5 \pm 4.1 $ & G  \\
01-50 & GLEAM J130459$-$210638 & 196.24867 & $ -21.11480 $ & $ 1.85 \pm 0.02 $ & $ 62.9 \pm 4.5 $ & $ 3.4 \pm 0.2 $ & $ -11.9 \pm 0.3 $ & $ 5.53 \pm 0.05 $ & $ -14.9 \pm 2.0 $ & G  \\
01-51 & GLEAM J132747$-$224152 & 201.94458 & $ -22.69700 $ & $ 0.80 \pm 0.01 $ & $ 36.8 \pm 8.0 $ & $ 4.6 \pm 1.0 $ & $ -34.9 \pm 0.8 $ & $ 8.35 \pm 0.17 $ & $ -34.8 \pm 4.7 $ & G  \\
01-52 & GLEAM J133007$-$214203 & 202.52727 & $ -21.69750 $ & $ 8.35 \pm 0.02 $ & $ 102.6 \pm 7.6 $ & $ 1.2 \pm 0.1 $ & $ -21.5 \pm 0.2 $ & $ 2.68 \pm 0.02 $ & $ -9.1 \pm 1.7 $ & G  \\
01-53 & GLEAM J133515$-$255216 & 203.80841 & $ -25.87420 $ & $ 1.14 \pm 0.01 $ & $ 31.4 \pm 4.0 $ & $ 2.8 \pm 0.3 $ & $ -14.7 \pm 0.3 $ & $ 5.31 \pm 0.10 $ & $ -22.6 \pm 3.6 $ & G  \\
01-54 & GLEAM J134038$-$340234 & 205.14667 & $ -34.04030 $ & $ 0.18 \pm 0.04 $ & $ 42.0 \pm 8.6 $ & $ 23.3 \pm 7.0 $ & $ -60.0 \pm 0.7 $ & $ - $ & $ - $ & G  \\
01-55 & GLEAM J134744$-$260255 & 206.92415 & $ -26.05270 $ & $ 0.74 \pm 0.01 $ & $ 28.0 \pm 5.0 $ & $ 3.8 \pm 0.7 $ & $ -19.2 \pm 0.5 $ & $ 8.22 \pm 0.14 $ & $ -16.6 \pm 3.8 $ & P  \\
01-56 & GLEAM J140217$-$230405 & 210.57090 & $ -23.06620 $ & $ 0.47 \pm 0.02 $ & $ 26.6 \pm 5.7 $ & $ 5.7 \pm 1.2 $ & $ -23.0 \pm 0.7 $ & $ 8.62 \pm 0.29 $ & $ -26.0 \pm 6.9 $ & G  \\
01-57 & GLEAM J140239$-$285003 & 210.65822 & $ -28.83680 $ & $ 0.79 \pm 0.01 $ & $ 47.4 \pm 4.5 $ & $ 6.0 \pm 0.6 $ & $ -28.0 \pm 0.3 $ & $ 10.22 \pm 0.17 $ & $ -30.1 \pm 3.5 $ & P  \\
01-58 & GLEAM J151540$-$193945 & 228.92467 & $ -19.66540 $ & $ 2.12 \pm 0.01 $ & $ 76.2 \pm 6.8 $ & $ 3.6 \pm 0.3 $ & $ -9.1 \pm 0.3 $ & $ 6.46 \pm 0.22 $ & $ -19.3 \pm 8.0 $ & P  \\
01-59 & GLEAM J151742$-$242216 & 229.42553 & $ -24.36500 $ & $ 1.84 \pm 0.01 $ & $ 26.2 \pm 5.6 $ & $ 1.4 \pm 0.3 $ & $ -11.5 \pm 0.8 $ & $ 5.67 \pm 0.02 $ & $ -5.4 \pm 0.9 $ & G  \\
01-60 & GLEAM J191024$-$290609 & 287.60679 & $ -29.09890 $ & $ 0.69 \pm 0.02 $ & $ 61.4 \pm 8.6 $ & $ 8.9 \pm 1.3 $ & $ -25.7 \pm 0.4 $ & $ 11.68 \pm 0.26 $ & $ -34.9 \pm 5.3 $ & G  \\
01-61 & GLEAM J191928$-$295756 & 289.86426 & $ -29.96060 $ & $ 6.20 \pm 0.02 $ & $ 54.6 \pm 11.3 $ & $ 0.9 \pm 0.2 $ & $ -31.4 \pm 0.7 $ & $ 1.47 \pm 0.03 $ & $ -43.5 \pm 5.5 $ & G  \\
01-62 & GLEAM J194554$-$270617 & 296.47192 & $ -27.10410 $ & $ 4.14 \pm 0.03 $ & $ 37.0 \pm 7.9 $ & $ 0.9 \pm 0.2 $ & $ -22.1 \pm 0.6 $ & $ 6.38 \pm 0.07 $ & $ -22.5 \pm 2.6 $ & G  \\
01-63 & GLEAM J194822$-$302010 & 297.09591 & $ -30.33710 $ & $ 2.45 \pm 0.01 $ & $ 43.3 \pm 7.7 $ & $ 1.8 \pm 0.3 $ & $ -31.4 \pm 0.7 $ & $ 9.74 \pm 0.21 $ & $ -42.3 \pm 3.3 $ & P  \\
01-64 & GLEAM J200224$-$271549 & 300.60983 & $ -27.25740 $ & $ 1.04 \pm 0.02 $ & $ 43.8 \pm 5.0 $ & $ 4.2 \pm 0.5 $ & $ -36.6 \pm 0.3 $ & $ 10.67 \pm 0.18 $ & $ -23.5 \pm 3.6 $ & P  \\
01-65 & GLEAM J201339$-$272417 & 303.41618 & $ -27.41270 $ & $ 1.75 \pm 0.02 $ & $ 48.3 \pm 4.7 $ & $ 2.8 \pm 0.3 $ & $ -13.7 \pm 0.3 $ & $ 8.15 \pm 0.29 $ & $ -6.4 \pm 7.9 $ & P  \\
01-66 & GLEAM J201450$-$222505 & 303.71250 & $ -22.41910 $ & $ 0.62 \pm 0.02 $ & $ 44.9 \pm 6.7 $ & $ 7.2 \pm 1.1 $ & $ -26.3 \pm 0.4 $ & $ 9.99 \pm 0.24 $ & $ -26.6 \pm 5.2 $ & P  \\
01-67 & GLEAM J201707$-$310305 & 304.31277 & $ -31.05510 $ & $ 0.92 \pm 0.01 $ & $ 45.1 \pm 7.8 $ & $ 4.9 \pm 0.8 $ & $ -23.3 \pm 0.6 $ & $ 19.06 \pm 0.38 $ & $ -19.4 \pm 3.4 $ & P  \\
01-68 & GLEAM J202803$-$315507 & 307.01840 & $ -31.91650 $ & $ 1.75 \pm 0.05 $ & $ 52.4 \pm 5.9 $ & $ 3.0 \pm 0.3 $ & $ -19.5 \pm 0.3 $ & $ 2.93 \pm 0.10 $ & $ -38.1 \pm 6.5 $ & G  \\
01-69 & GLEAM J203316$-$225314 & 308.32698 & $ -22.88150 $ & $ 8.88 \pm 0.02 $ & $ 104.1 \pm 7.7 $ & $ 1.2 \pm 0.1 $ & $ -26.0 \pm 0.3 $ & $ 6.69 \pm 0.05 $ & $ -25.6 \pm 0.9 $ & G  \\
01-70 & GLEAM J205206$-$282913 & 313.02914 & $ -28.48780 $ & $ 1.24 \pm 0.01 $ & $ 29.1 \pm 6.8 $ & $ 2.4 \pm 0.5 $ & $ 11.8 \pm 0.7 $ & $ - $ & $ - $ & G  \\
01-71 & GLEAM J211406$-$220705 & 318.52307 & $ -22.12180 $ & $ 0.97 \pm 0.02 $ & $ 36.0 \pm 4.2 $ & $ 3.7 \pm 0.4 $ & $ -14.9 \pm 0.4 $ & $ - $ & $ - $ & G  \\
01-72 & GLEAM J212232$-$230151 & 320.64368 & $ -23.03140 $ & $ 0.55 \pm 0.03 $ & $ 34.3 \pm 5.7 $ & $ 6.2 \pm 1.1 $ & $ -12.4 \pm 0.5 $ & $ - $ & $ - $ & P  \\
01-73 & GLEAM J215506$-$321945 & 328.77505 & $ -32.32570 $ & $ 2.40 \pm 0.01 $ & $ 75.4 \pm 5.8 $ & $ 3.1 \pm 0.2 $ & $ 20.8 \pm 0.3 $ & $ 5.08 \pm 0.30 $ & $ 17.0 \pm 9.5 $ & P  \\
01-74 & GLEAM J215611$-$352553 & 329.05010 & $ -35.43380 $ & $ 2.45 \pm 0.02 $ & $ 79.9 \pm 16.1 $ & $ 3.3 \pm 0.7 $ & $ 29.8 \pm 0.5 $ & $ 7.78 \pm 0.10 $ & $ 25.9 \pm 2.0 $ & G*  \\
01-75 & GLEAM J215657$-$181343 & 329.22708 & $ -18.22270 $ & $ 5.39 \pm 0.02 $ & $ 110.3 \pm 7.3 $ & $ 2.0 \pm 0.1 $ & $ 15.8 \pm 0.3 $ & $ 4.48 \pm 0.07 $ & $ 24.5 \pm 3.3 $ & P  \\
01-76 & GLEAM J223919$-$261008 & 339.83815 & $ -26.16530 $ & $ 1.67 \pm 0.01 $ & $ 41.7 \pm 2.9 $ & $ 2.5 \pm 0.2 $ & $ 12.0 \pm 0.4 $ & $ 2.58 \pm 0.08 $ & $ 13.2 \pm 6.8 $ & G  \\
01-77 & GLEAM J224647$-$281746 & 341.69838 & $ -28.30060 $ & $ 0.75 \pm 0.01 $ & $ 23.5 \pm 3.4 $ & $ 3.1 \pm 0.5 $ & $ 11.0 \pm 0.7 $ & $ - $ & $ - $ & G*  \\
01-78 & GLEAM J225104$-$220431 & 342.77560 & $ -22.07230 $ & $ 1.88 \pm 0.01 $ & $ 44.2 \pm 4.2 $ & $ 2.4 \pm 0.2 $ & $ 17.4 \pm 0.5 $ & $ 5.63 \pm 0.11 $ & $ 18.0 \pm 4.7 $ & G  \\
01-79 & GLEAM J231555$-$282615 & 348.98088 & $ -28.43180 $ & $ 0.79 \pm 0.02 $ & $ 45.4 \pm 3.6 $ & $ 5.7 \pm 0.5 $ & $ 14.9 \pm 0.3 $ & $ - $ & $ - $ & G  \\
01-80 & GLEAM J234945$-$292024 & 357.45095 & $ -29.34000 $ & $ 1.27 \pm 0.02 $ & $ 52.4 \pm 4.3 $ & $ 4.1 \pm 0.3 $ & $ 14.5 \pm 0.3 $ & $ 12.92 \pm 0.13 $ & $ 17.3 \pm 2.1 $ & G  \\
01-81 & GLEAM J235137$-$300751 & 357.91401 & $ -30.12600 $ & $ 0.85 \pm 0.01 $ & $ 26.3 \pm 8.5 $ & $ 3.1 \pm 1.0 $ & $ 13.4 \pm 1.6 $ & $ 1.78 \pm 0.13 $ & $ 29.5 \pm 15.3 $ & G  
\end{longtable}
}
\end{landscape}

{\footnotesize
\begin{longtable}{p{1cm}lccp{7cm}}
\caption{Classifications for sources detected in the Dec$-27\degree$ strip.}\label{tab:class}\\
\hline
POGS ID & GLEAM ID & Morphology$^{\ast}$ & Host$^{\dag}$ & Notes \& References$^{\ddag}$ \\
\hline
\endfirsthead
\caption{continued.}\\
\hline
POGS ID & GLEAM ID & Morphology$^{\ast}$ & Host$^{\dag}$ & Notes \& References$^{\ddag}$ \\
\hline
\endhead
\hline
\multicolumn{5}{l}{$^{\ast}$CS: compact-single. CD: compact-double. ED: extended-double. $^{\dag}$EG: elliptical galaxy. PSR: pulsar. ?: host could not be }\\
\multicolumn{5}{l}{identified. $^{\ddag}$All redshifts are from \cite{Flesch2017} unless otherwise specified.}\\
\endfoot
\hline
\multicolumn{5}{l}{$^{\ast}$CS: compact-single. CD: compact-double. ED: extended-double. $^{\dag}$?: host could not be identified. $^{\ddag}$All redshifts are from}\\
\multicolumn{5}{l}{\cite{Flesch2017} unless otherwise specified.}\\
\endlastfoot

\hline%
01-01 & GLEAM J001341$-$300925 & CS \hspace{3mm} & AGN & $z=1.111$ \\
01-02 & GLEAM J002021$-$202846 & CD \hspace{3mm} & AGN & $z=0.545$ \\
01-03 & GLEAM J002026$-$201427 & ED \hspace{3mm} & AGN & MRC~0017$-$205 \citep[$z =0.197$;][]{McCarthy1996} \\
01-04 & GLEAM J002112$-$191041 & ED \hspace{3mm} & AGN & PKS~0018$-$19 \citep[$z = 0.096$;][]{Jones2009} \\
01-05 & GLEAM J003906$-$242506 & CD \hspace{3mm} & ? & \\
01-06 & GLEAM J014709$-$223234 & CS \hspace{3mm} & AGN & MRC~0144$-$227 \citep[$z =0.6$;][]{McCarthy1996} \\
01-07 & GLEAM J021726$-$224207 & CS \hspace{3mm} & ? & \\
01-08 & GLEAM J023512$-$293622 & CD \hspace{3mm} & AGN & $z=0.0599$ \citep{Jones2009} \\
01-09 & GLEAM J023700$-$283448 & CD \hspace{3mm} & AGN & $z=0.1447$ \citep{Sadler2002} \\
01-10 & GLEAM J023845$-$223321 & CD \hspace{3mm} & AGN & $z_{\rm{phot}}=0.5$ \\
01-11 & GLEAM J031152$-$312959 & CD \hspace{3mm} & AGN & $z=2.417$ \\
01-12 & GLEAM J032022$-$312458 & CS \hspace{3mm} & ? &  \\
01-13 & GLEAM J032800$-$220205 & CD \hspace{3mm} & AGN & $z=2.2$ \\
01-14 & GLEAM J033810$-$250213 & CS \hspace{3mm} & AGN & $z_{\rm{phot}}=0.6$ \\
01-15 & GLEAM J034850$-$294359 & CD \hspace{3mm} & AGN & Unknown $z$ \citep{McCarthy1996} \\
01-16 & GLEAM J035125$-$274610 & ED \rdelim\}{2}{3mm} & \multirow{2}{*}{Seyfert II} & \multirow{2}{7cm}{SE \& NW lobes of PMN~J0351$-$2744 \citep[$z=0.0656$;][]{Mahony2011}} \\
01-17 & GLEAM J035140$-$274354 & ED \hspace{3mm} & & \\
01-18 & GLEAM J035537$-$203324 & CS \hspace{3mm} & AGN & Unknown $z$ \citep{McCarthy1996} \\
01-19 & GLEAM J044119$-$295231 & CS \hspace{3mm} & AGN & Unknown $z$ \\
01-20 & GLEAM J045714$-$270436 & CS \hspace{3mm} & ? & \\
01-21 & GLEAM J050535$-$285648 & ED \rdelim\}{2}{3mm} & \multirow{2}{*}{LINER} & \multirow{2}{7cm}{S lobe backflow and N lobe of ESO~422$-$G028 \citep[$z=0.0381$;][]{Jamrozy2005}} \\ 
01-22 & GLEAM J050544$-$282236 & ED \hspace{3mm} & & \\
01-23 & GLEAM J050922$-$282427 & CS \hspace{3mm} & AGN & $z=0.06$ \citep{Jones2009} \\
01-24 & GLEAM J051024$-$195950 & CS \hspace{3mm} & AGN & $z_{\rm{phot}}=0.6$ \\
01-25 & GLEAM J051836$-$240305 & CD \hspace{3mm} & AGN & $z=0.0339$ \citep{Jones2009} \\
01-26 & GLEAM J052618$-$260409 & CS \hspace{3mm} & ? & \\
01-27 & GLEAM J055222$-$261015 & CS \hspace{3mm} & ? & \\
01-28 & GLEAM J060753$-$263806 & CD \hspace{3mm} & ? &  \\
01-29 & GLEAM J060845$-$271708 & CS \hspace{3mm} & AGN & $z_{\rm{phot}}=0.8$ \\
01-30 & GLEAM J062126$-$210059 & CS \hspace{3mm} & AGN & $z_{\rm{phot}}=1.3$ \\
01-31 & GLEAM J062702$-$241925 & CS \hspace{3mm} & AGN & $z_{\rm{phot}}=0.9$ \\
01-32 & GLEAM J063049$-$283438 & CS \hspace{3mm} & PSR & PSR~B0630$-2834$ \citep{Johnston2008}\\
01-33 & GLEAM J063228$-$272109 & CD \hspace{3mm} & ? & \\
01-34 & GLEAM J063631$-$202924 & ED \rdelim\}{2}{3mm} & \multirow{2}{*}{EG} & \multirow{2}{7cm}{N and S hotspot of PKS~J0636$-2036$ \citep[$z=0.0551$;][]{McAdam1975}} \\
01-35 & GLEAM J063633$-$204225 & ED \hspace{3mm} & & \\
01-36 & GLEAM J065716$-$320328 & CS \hspace{3mm} & AGN & $z_{\rm{phot}}=1.3$ \\
01-37 & GLEAM J070901$-$355921 & ED \rdelim\}{2}{3mm} & \multirow{2}{*}{Seyfert II} & \multirow{2}{7cm}{NE and SW hotspots of PKS~0707$-35$ \citep[$z=0.111$;][]{Burgess2006}}\\
01-38 & GLEAM J070934$-$360341 & ED \hspace{3mm} & & \\
01-39 & GLEAM J091245$-$251254 & CD \hspace{3mm} & AGN &  $z_{\rm{phot}}=0.2$ \\
01-40 & GLEAM J092317$-$213744 & CD \hspace{3mm} & ? & \\
01-41 & GLEAM J095750$-$283808 & CD \hspace{3mm} & AGN & $z=0.346$ \citep{KV09} \\
01-42 & GLEAM J100055$-$193108 & CD \hspace{3mm} & AGN & $z_{\rm{phot}}=1.2$ \\
01-43 & GLEAM J100206$-$265606 & CS \hspace{3mm} & AGN & $z_{\rm{phot}}=0.9$ \\
01-44 & GLEAM J102056$-$321100 & CD \hspace{3mm} & AGN & $z_{\rm{phot}}=1.2$ \\
01-45 & GLEAM J110331$-$325116 & CD \hspace{3mm} & AGN & $z=0.35$ \\
01-46 & GLEAM J111007$-$272960 & CS \hspace{3mm} & AGN & Unknown $z$ \citep{McCarthy1996} \\
01-47 & GLEAM J120533$-$263407 & CS \hspace{3mm} & AGN & $z=0.786$ \\
01-48 & GLEAM J122646$-$324711 & CS \hspace{3mm} & ? & \\
01-49 & GLEAM J130025$-$231806 & CD \hspace{3mm} & AGN & $z=1.109$ \\
01-50 & GLEAM J130459$-$210638 & CS \hspace{3mm} & AGN & Unknown $z$ \citep{Souchay2015} \\
01-51 & GLEAM J132747$-$224152 & CS \hspace{3mm} & ? & \\
01-52 & GLEAM J133007$-$214203 & CS \hspace{3mm} & AGN & $z=0.525$ \\
01-53 & GLEAM J133515$-$255216 & CS \hspace{3mm} & ? & \\
01-54 & GLEAM J134038$-$340234 & CS \hspace{3mm} & ? & \\
01-55 & GLEAM J134744$-$260255 & CD \hspace{3mm} & ? & \\
01-56 & GLEAM J140217$-$230405 & CS \hspace{3mm} & ? & \\
01-57 & GLEAM J140239$-$285003 & CD \hspace{3mm} & ? & \\
01-58 & GLEAM J151540$-$193945 & CD \hspace{3mm} & ? & \\
01-59 & GLEAM J151742$-$242216 & CS \hspace{3mm} & AGN & BL~Lac at $z=0.048$ \\
01-60 & GLEAM J191024$-$290609 & CD \hspace{3mm} & ? & \\
01-61 & GLEAM J191928$-$295756 & CS \hspace{3mm} & AGN & $z=0.167$ \citep{Jones2009} \\
01-62 & GLEAM J194554$-$270617 & CS \hspace{3mm} & ? & \\
01-63 & GLEAM J194822$-$302010 & CD \hspace{3mm} & ? & \\
01-64 & GLEAM J200224$-$271549 & CD \hspace{3mm} & AGN & $z_{\rm{phot}}=0.30$ \\
01-65 & GLEAM J201339$-$272417 & CD \hspace{3mm} & ? & \\
01-66 & GLEAM J201450$-$222505 & CD \hspace{3mm} & ? & \\
01-67 & GLEAM J201707$-$310305 & CD \hspace{3mm} & ? & This source is comprised of a pair of CD sources. \\
01-68 & GLEAM J202803$-$315507 & CS \hspace{3mm} & AGN & $z_{\rm{phot}}=1.00$ \\
01-69 & GLEAM J203316$-$225314 & CS \hspace{3mm} & AGN & $z=0.131$ \citep{Jones2009} \\
01-70 & GLEAM J205206$-$282913 & CS \hspace{3mm} & ? & \\
01-71 & GLEAM J211406$-$220705 & CD \hspace{3mm} & ? & \\
01-72 & GLEAM J212232$-$230151 & CD \hspace{3mm} & ? & \\
01-73 & GLEAM J215506$-$321945 & CD \hspace{3mm} & ? & \\
01-74 & GLEAM J215611$-$352553 & CD \hspace{3mm} & AGN & $z_{\rm{phot}}=0.9$ \\
01-75 & GLEAM J215657$-$181343 & CD \hspace{3mm} & AGN & $z=0.668$ \\
01-76 & GLEAM J223919$-$261008 & CS \hspace{3mm} & ? & \\
01-77 & GLEAM J224647$-$281746 & CD \hspace{3mm} & ? & \\
01-78 & GLEAM J225104$-$220431 & CS \hspace{3mm} & AGN & $z=0.741$ \citep{Jones2009} \\
01-79 & GLEAM J231555$-$282615 & ED \hspace{3mm} & AGN & $z=0.2293$ \citep{Colless2001} \\
01-80 & GLEAM J234945$-$292024 & CD \hspace{3mm} & AGN & $z=0.223$ \citep{Colless2001} \\
01-81 & GLEAM J235137$-$300751 & CS \hspace{3mm} & AGN & Unknown $z$ \citep{Mahony2011} \\
\hline\hline
\end{longtable}
}

\begin{figure*}
\begin{center}
\includegraphics[width=0.9\textwidth]{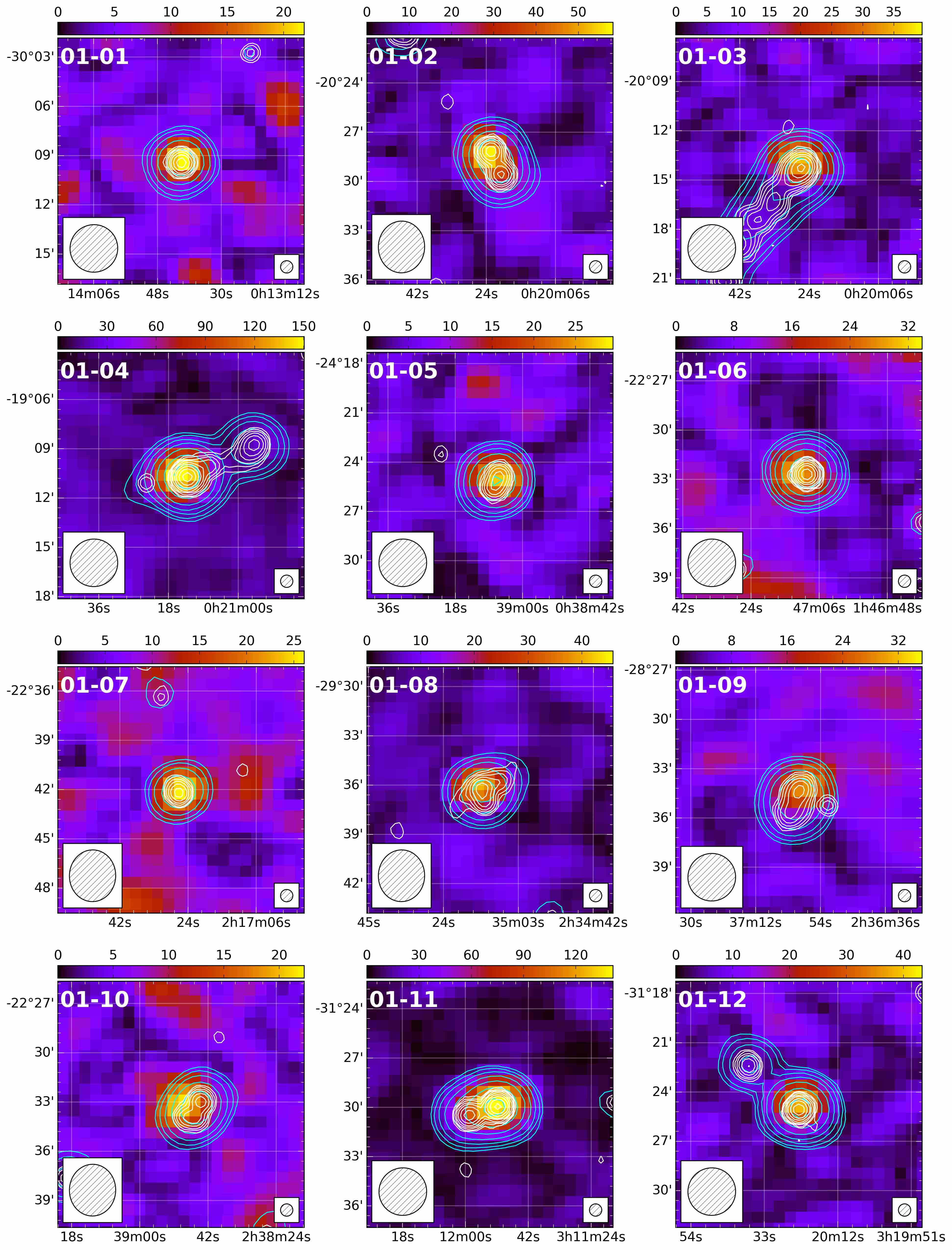}
\caption{Postage stamp images of POGS sources. Each panel shows a slice through our FD cubes near the fitted source FD (see Table~\ref{tab:catalogue}). Cyan contours are continuum from the GLEAM "white" mosaic (204--232~MHz) starting at $5\sigma_{\rm{local}}$ and scaling by a factor two, where $\sigma_{\rm{local}}$ is the `wide-band' rms \citep[see][]{HurleyWalker2017}. White contours are NVSS continuum, starting at 2.25 mJy PSF$^{-1}$ $(5\sigma)$ and scaling by a factor two. Right Ascension and Declination are in J2000 coordinates; color scale is in mJy PSF$^{-1}$. The ID number in the upper-left corner corresponds to the source ID in Table~\ref{tab:catalogue} and Table~\ref{tab:class}. Each subplot shows a $15\times15$~arcmin region.}\label{fig:multi_postage}
\end{center}
\end{figure*}

\begin{figure*}
\ContinuedFloat
\begin{center}
\includegraphics[width=0.9\textwidth]{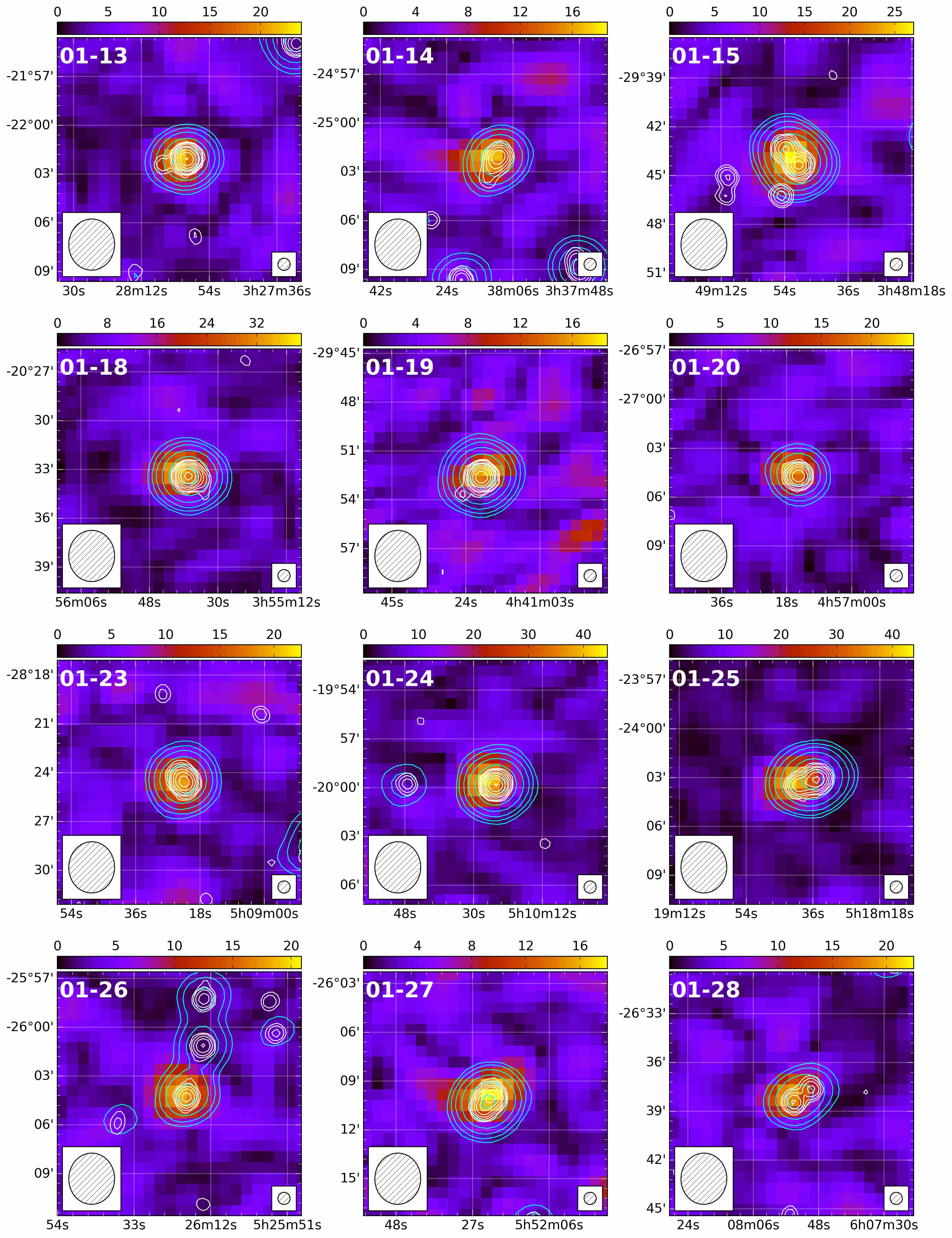}
\caption{(Continued from previous page)}\label{fig:multi_postage}
\end{center}
\end{figure*}

\begin{figure*}
\ContinuedFloat
\begin{center}
\includegraphics[width=0.9\textwidth]{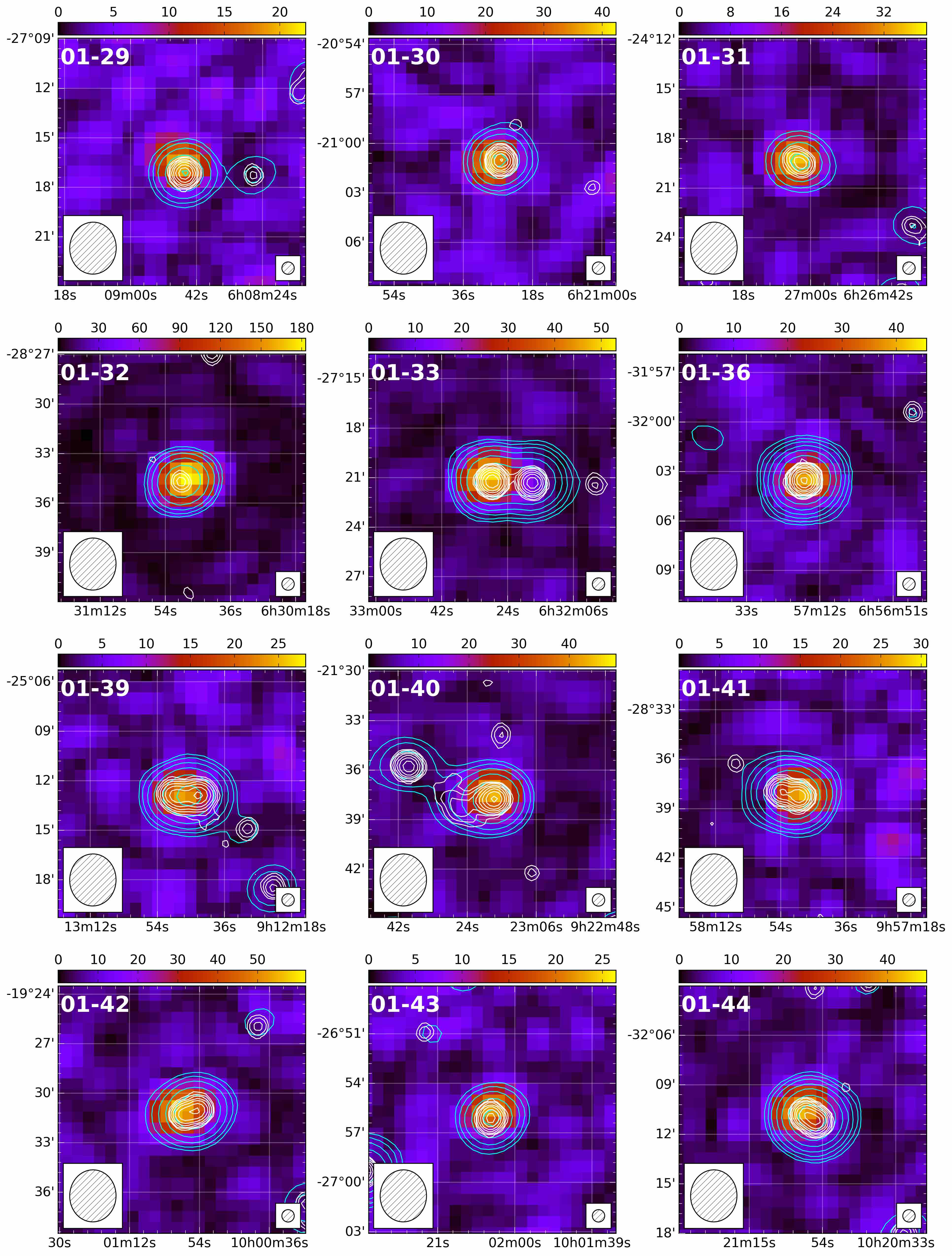}
\caption{(Continued from previous page)}\label{fig:multi_postage}
\end{center}
\end{figure*}

\begin{figure*}
\ContinuedFloat
\begin{center}
\includegraphics[width=0.9\textwidth]{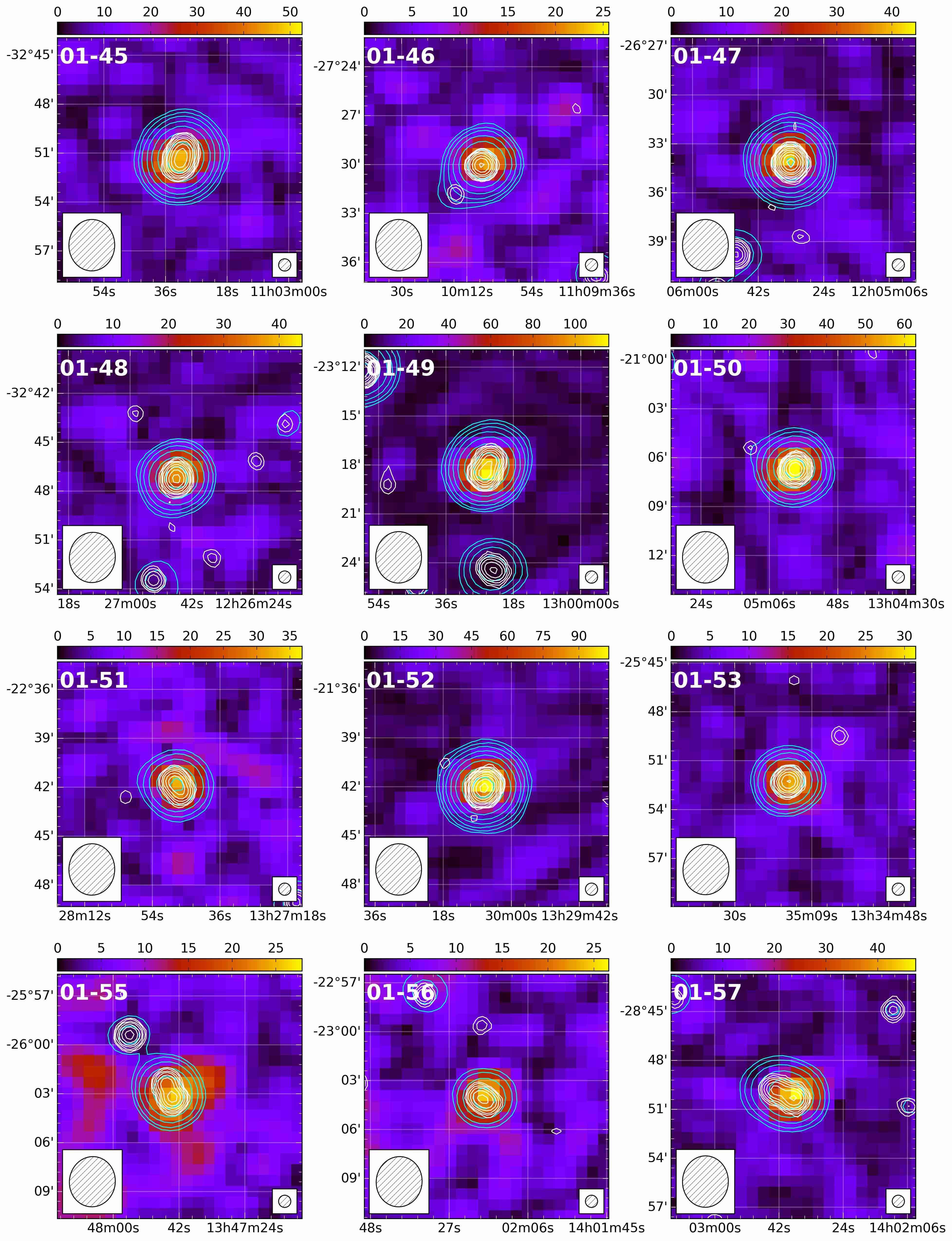}
\caption{(Continued from previous page)}\label{fig:multi_postage}
\end{center}
\end{figure*}

\begin{figure*}
\ContinuedFloat
\begin{center}
\includegraphics[width=0.9\textwidth]{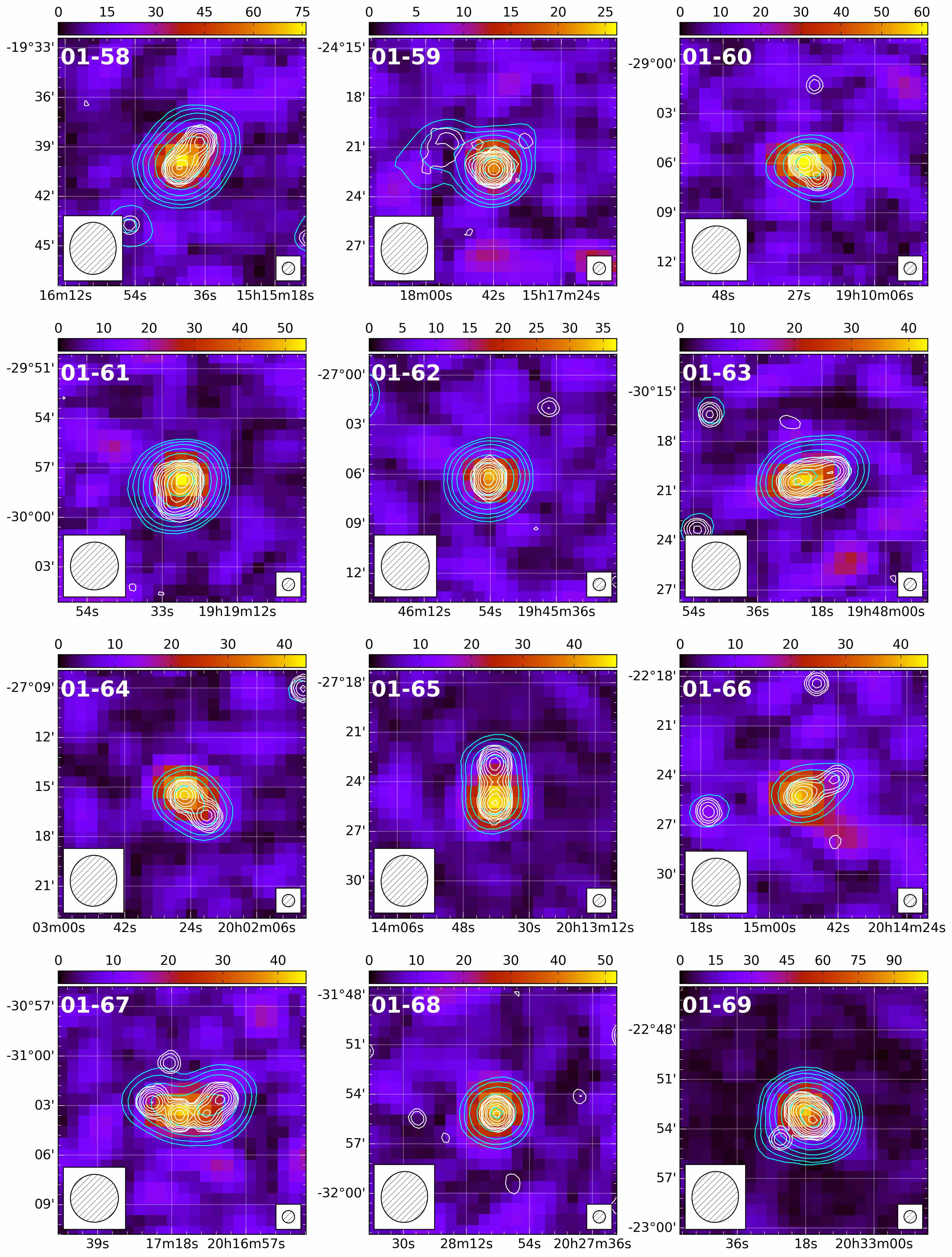}
\caption{(Continued from previous page)}\label{fig:multi_postage}
\end{center}
\end{figure*}

\begin{figure*}
\ContinuedFloat
\begin{center}
\includegraphics[width=0.9\textwidth]{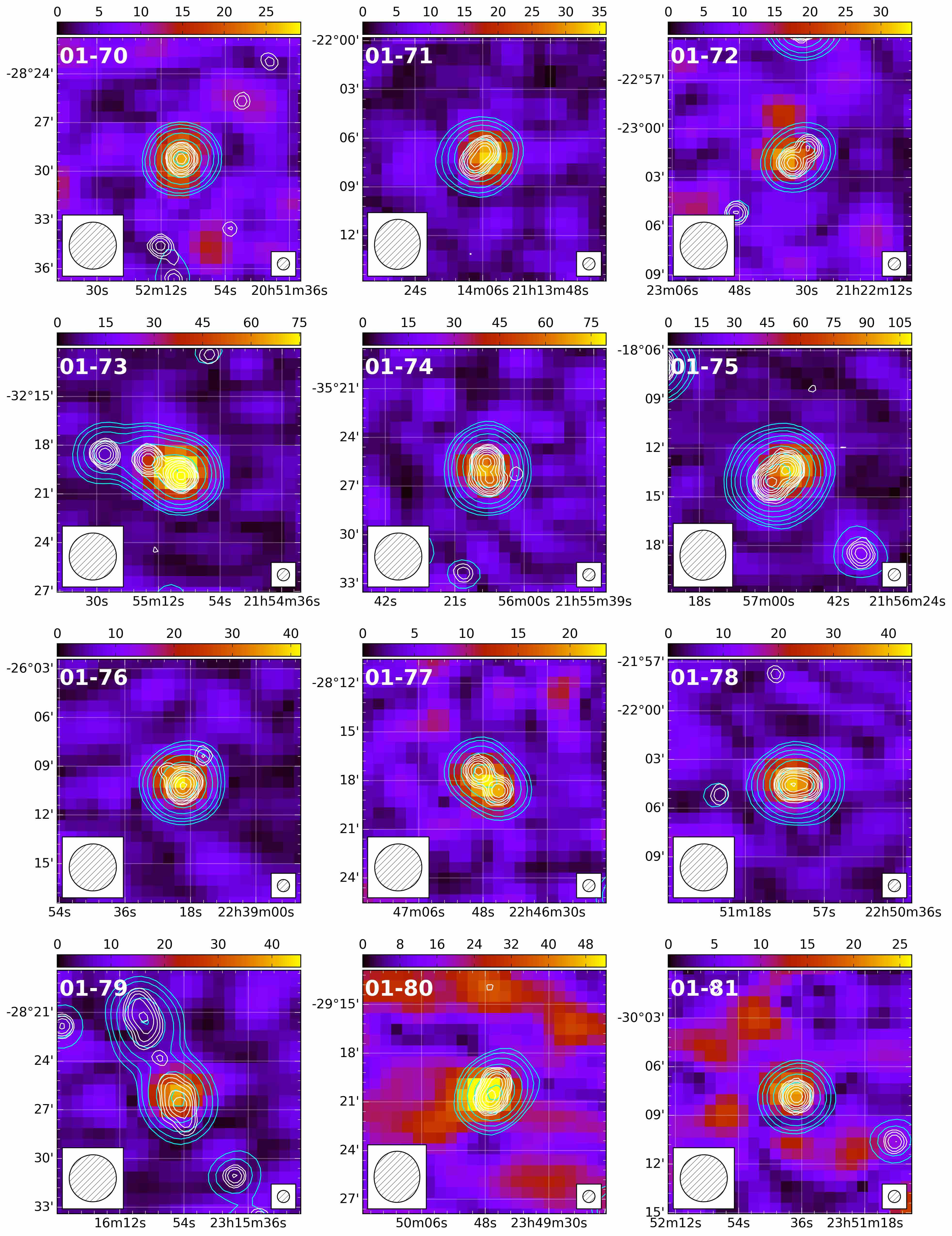}
\caption{(Continued from previous page)}\label{fig:multi_postage}
\end{center}
\end{figure*}

\end{document}